\documentclass[useAMS]{mn2e}

\usepackage{psfig}

\def\zabs{$z_{\rm abs}$}

\def\lya{Lyman-$\alpha$ }

\def\h2{H$_2$}
\def\hi{H~{\sc i}~}

\def\kms{km~s$^{-1}$}
\def\nh{n$_{\rm H}$~}
\def\ltsima{$\; \buildrel < \over \sim \;$}
\def\simlt{\lower.5ex\hbox{\ltsima}}
\def\gtsima{$\; \buildrel > \over \sim \;$}
\def\simgt{\lower.5ex\hbox{\gtsima}}

\begin{document}

\title[UVES-VLT survey for H$_2$ in DLA systems]{The VLT-UVES survey for molecular hydrogen in 
high-redshift damped Lyman-$\alpha$  systems: Physical conditions in the neutral gas 
}

\author[Srianand et al.]{R. Srianand$^1$, 
Patrick Petitjean$^{2,3}$,  C\'edric Ledoux$^4$, Gary Ferland$^5$, Gargi Shaw$^5$\\
$^1$ IUCAA, Post Bag 4, Ganesh Khind, Pune 411 007, India - email: anand@iucaa.ernet.in\\
$^2$ Institut d'Astrophysique de Paris -- CNRS, 98bis Boulevard Arago, F-75014 Paris, France - email: petitjean@iap.fr\\
$^3$ LERMA, Observatoire de Paris, 61 Avenue de l'Observatoire, F-75014, Paris, France\\
$^4$ European Southern Observatory, Alonso de C\'ordova 3107, Casilla 19001, Vitacura, Santiago, Chile - email: cledoux@eso.org\\
$^5$ Department of Physics and Astronomy, University of
Kentucky, 177 Chemistry/Physics Building, Lexington, KY 40506\\
~~~email: gary@pa.uky.edu and gargi@pa.uky.edu\\}

\date{Received date / Accepted date}
\pubyear{2004} \volume{000} \pagerange{1}

\maketitle 

\begin{abstract}
{
We study the physical conditions in damped Lyman-$\alpha$ systems (DLAs), 
using a sample of 33 systems toward 26 QSOs acquired for a
recently completed survey of \h2 by Ledoux et al. (2003). 
We use the column densities of \h2 in different rotational levels, 
together with those of C~{\sc i}, C~{\sc i}$^*$, C~{\sc i}$^{**}$, 
C~{\sc ii}$^*$ and singly ionized atomic species to discuss the kinetic 
temperature, the density of hydrogen and the electronic density in the 
gas together with the ambient UV radiation field. Detailed comparisons 
are made between the observed properties in DLAs, the interstellar 
medium (ISM) of the Galaxy, the large and small Magellanic 
clouds (LMC and SMC).  

\par
The mean kinetic temperature of the gas corresponding to DLA 
subcomponents in which \h2 absorption line is detected, derived 
from the ortho-to-para ratio (153$\pm$78 K),
is higher than that measured in the ISM (77$\pm17$ K)
and the Magellanic clouds (82$\pm$21 K). Typical pressure in these components
(corresponding to $T$~=~100$-$300 K and $n_{\rm H}$ = 10$-$200 cm$^{-3}$), 
measured using C~{\sc i} 
fine-structure excitation, are higher than what is measured along ISM 
sightlines. This is consistent with the corresponding 
higher values for $N$(\h2,J=2)/$N$(\h2,J=0) seen in DLAs. 
From the column densities of the high-J rotational levels, 
we derive that the typical radiation field in the \h2 bearing components is 
of the order of or slightly higher than the mean UV field in the Galactic 
ISM.  Determination of electron density in the gas with \h2
and C~{\sc i} show the ionization rate is similar to that 
of a cold neutral medium (CNM) in a moderate radiation field.
This, together with the fact that we see \h2 in 13-20\% of the DLAs,
can be used to conclude that DLAs at $z>1.9$ could contribute
as much as 50\% star formation rate density seen in Lyman
break galaxies (LBGs).

\par
C~{\sc ii$^*$} absorption line is detected in all the components where \h2
absorption line is seen. The excitation of C~{\sc ii} in these systems is consistent 
with the physical parameters derived from the excitation of \h2 and C~{\sc i}. 
We detect C~{\sc ii$^*$} in about 50\% of the DLAs and therefore in 
a considerable fraction of DLAs that do not show \h2. 
In part of the later systems, 
physical conditions could be similar to that in the CNM gas of the Galaxy. 
However, the absence of C~{\sc i} absorption line
and the presence of Al~{\sc iii} absorption lines with a profile similar 
to the profiles of singly ionized species suggest an appreciable contribution from
warm (WNM) and/or partially ionized gas. The absence
of \h2, for the level of metallicity and dust depletion seen in these
systems, are consistent with low densities (i.e \nh$\le$ 1 cm$^{-3}$)
for a radiation field similar to the mean Galactic UV field.
}
\end{abstract}

\section{Introduction}
Damped Ly-$\alpha$ (DLA) systems seen in
QSO spectra are characterized by very large neutral hydrogen column densities:
$N($H~{\sc i}$)\ga 2\times 10^{20}$ cm$^{-2}$. Such an amount of neutral gas is 
usually  measured through local spiral disks. The case
for DLA systems to arise through proto-galactic disks is further supported by
the fact that the cosmological density of the absorbing gas
at $z_{\rm abs}\sim 3$ is of the same order of magnitude as the cosmological
density of stars at present epochs (Wolfe 1995).
Moreover, the presence of heavy elements ($Z\sim 1/10Z_\odot$) suggests that
DLAs are located in over-dense regions where star formation activity takes
place (Pettini et al. 1997) and at low and intermediate redshifts strong 
metal line systems and DLAs have been demonstrated to be associated with 
galaxies (e.g. Bergeron \& Boiss\'e, 1991; Le Brun et al. 1997).
It has also been shown that the profiles of the lines arising
in the neutral gas show evidence for rotation (e.g. Prochaska \& Wolfe 1997). 
However, hydrodynamical simulations have shown that the high redshift 
progenitors of present-day galactic disks could look like an aggregate 
of well separated dense clumps. In fact, the kinematics seen in the
absorption line profiles of DLAs could be explained by relative motions of the
clumps with little rotation (Haehnelt et al. 1998; Ledoux et al. 1998).
\par\noindent
Studying the star-formation activities in DLAs is very important 
for the understanding of galaxy formation in the Universe.
Recently, Wolfe et al. (2003a, 2003b, 2004) have shown that, even if DLAs 
sustain only a moderate star-formation activity, they will contribute appreciably 
to the global star-formation rate (SFR) density at high redshifts.
The SFR in DLAs can be estimated either by detecting the galaxies
responsible for DLAs or by inferring the intensity of the UV field in DLAs using
the induced excitation of atomic and molecular species. In the latter case, 
it is important to have a clear understanding of the physical conditions in the gas
to derive an accurate estimate of the SFR. In the case of the
Galactic ISM, rotational excitations of \h2 (see Browning et al. 2002
and references there in) and fine-structure excitations of C~{\sc i},
C~{\sc ii}, O~{\sc i} and Si~{\sc ii} are used to derive the physical
state of the absorbing gas (see for example Welty et al. 1999).
Detecting and studying these transitions in DLAs is the first
step toward understanding the physical conditions and hence the
star-formation activity in DLAs.
\par\noindent
Molecular hydrogen is ubiquitous in the neutral phase of the interstellar
medium (ISM) of galaxies. 
Formation of H$_2$ is expected on the surface of dust grains, 
if the gas is cool, dense and mostly neutral, and from the 
formation of H$^-$ ions if the gas is warm and dust-free
(see e.g. Jenkins \& Peimbert 1997; Cazaux \& Tielens 2002). 
As the former
process is most likely dominant in the neutral gas associated with DLA
systems, it is possible to obtain an indirect indication of the dust
content in DLAs without depending on extinction and/or
heavy element depletion effects. 
Moreover, by determining the populations of
different H$_2$ rotational levels, it is possible to constrain kinetic
and rotational excitation temperatures and particle densities. 
Effective photo-dissociation of H$_2$ takes place in the energy range
$11.1-13.6$ eV through Lyman- and Werner-band absorption lines and the
intensity of the local UV radiation field can therefore be derived from
the observed molecular fraction. A direct determination of the local
radiation field could have important implications in bridging the link
between DLA systems and star-formation activity at high redshifts.
\par\noindent
We have searched for molecular hydrogen in DLA and
sub-DLA systems at high redshift ($z_{\rm abs}>1.8$), using UVES at the VLT
down to a detection limit of typically $N($H$_2)\sim 2\times 10^{14}$
cm$^{-2}$ (see Ledoux et al. 2003). Out of the 33 systems in our sample, 
8 have firm and 2 have tentative detections of associated H$_2$ absorption lines.
In all of the systems, we measured metallicities relative to
Solar, [X/H] (with either X$=$Zn, or S, or Si), and depletion factors of iron,
[X/Fe], supposedly onto dust grains.
Although H$_2$ molecules are detected in systems with depletion
factor, [Zn/Fe], as low as 0.3, the systems where H$_2$ is detected are usually 
amongst those with the
highest metallicities and depletion factors. In particular, H$_2$ is
detected in the three systems with the largest depletion factors. Moreover, in
two different systems, one of the H$_2$-detected components has 
[Zn/Fe$]>1.5$.
This directly demonstrates that a large amount of dust is present in the
components where H$_2$ is detected.
The mean H$_2$ molecular fraction, $f=2N($H$_2)/[2N($H$_2)+N($H\,{\sc i}$)]$, in
DLA systems is generally small (typically $\log f<-1$) and similar to what is 
observed in the Magellanic Clouds.
There is no correlation between the amount of
molecules and the neutral hydrogen column density; in particular, two
systems where H$_2$ is detected have $\log N($H\,{\sc i}$)<20.3$.
Approximately 50 percent of the systems have $\log f<-6$: this is probably
a consequence of a reduced formation rate of H$_2$ onto dust grains 
(probably because the gas is warm, $T>1000$ K) and/or of an enhanced
ionizing flux relative to what is observed in our Galaxy. 
\par\noindent
In this work, we present additional high S/N ratio data on
three of the DLA systems in which \h2 is detected and the results of
multi-component Voigt profile fits to neutral and singly ionized species 
(including C~{\sc i}, C~{\sc i$^*$} and C~{\sc ii$^*$}) in all the 
DLAs in our sample.
We estimate the range of physical conditions in the neutral gas
using standard techniques that are used in ISM studies. The 
paper is organized as follows. In Section 2,
we give the details of the additional data and present the 
new fits to the \h2 absorption lines in the corresponding 
three systems. In Section 3,
we discuss the relative populations of different \h2 rotational levels 
deriving information on the physical state of the gas by
comparing the DLA observations with Galactic ISM, SMC, and LMC 
data. In Sections 4 and 5 we discuss, respectively, the fine-structure excitation
of C~{\sc i} and the ionization state of Carbon. In Section 6
we study the C~{\sc ii$^*$} excitation in detail. Finally,
we summarize our results and discuss various implications of
the overall study in Section 7.
\section{Data sample}
\begin{figure*}
\centerline{\vbox{
\psfig{figure=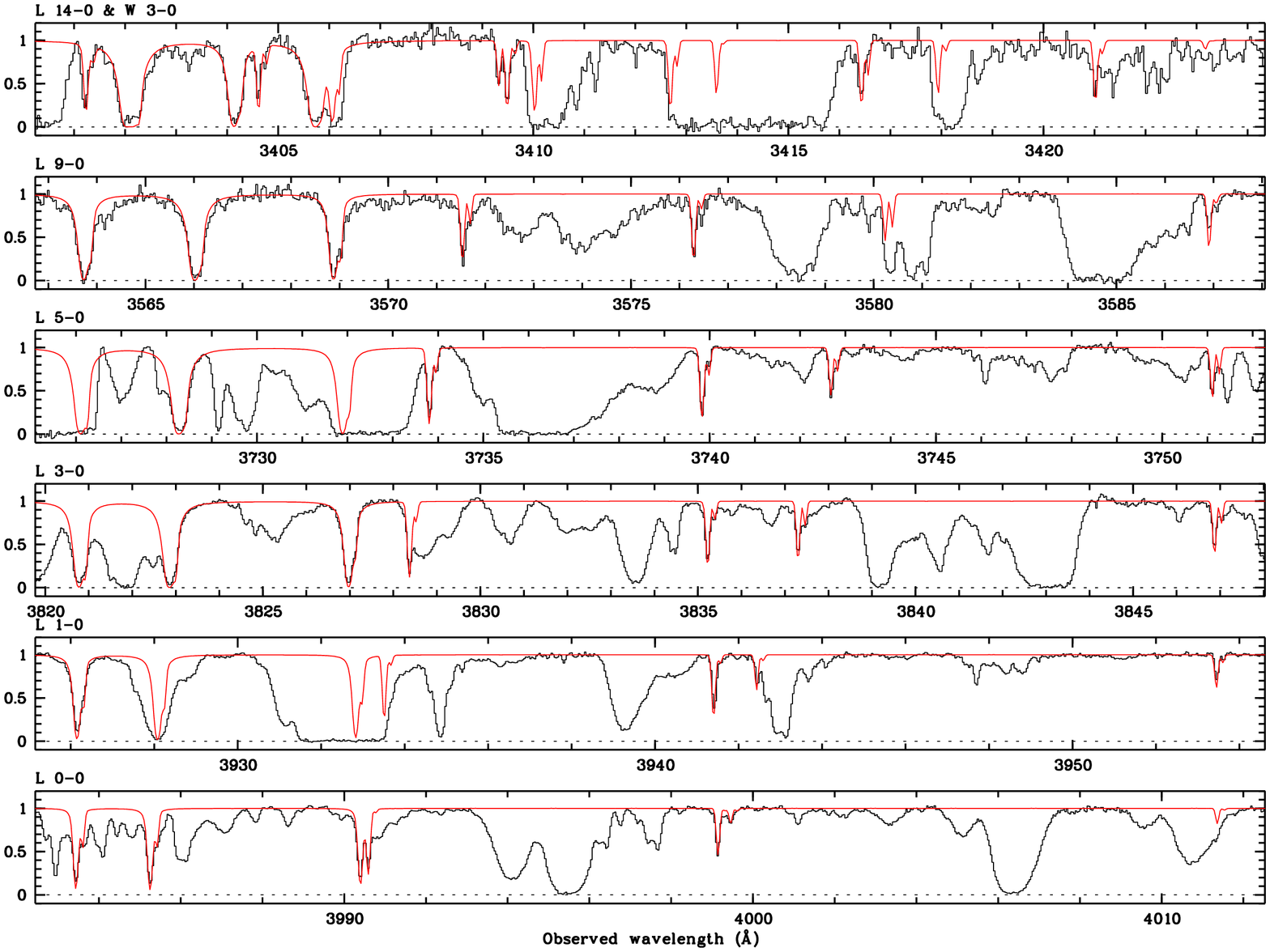,width=18.cm,height=10.cm,angle=270.}}}
\centerline{\vbox{
\psfig{figure=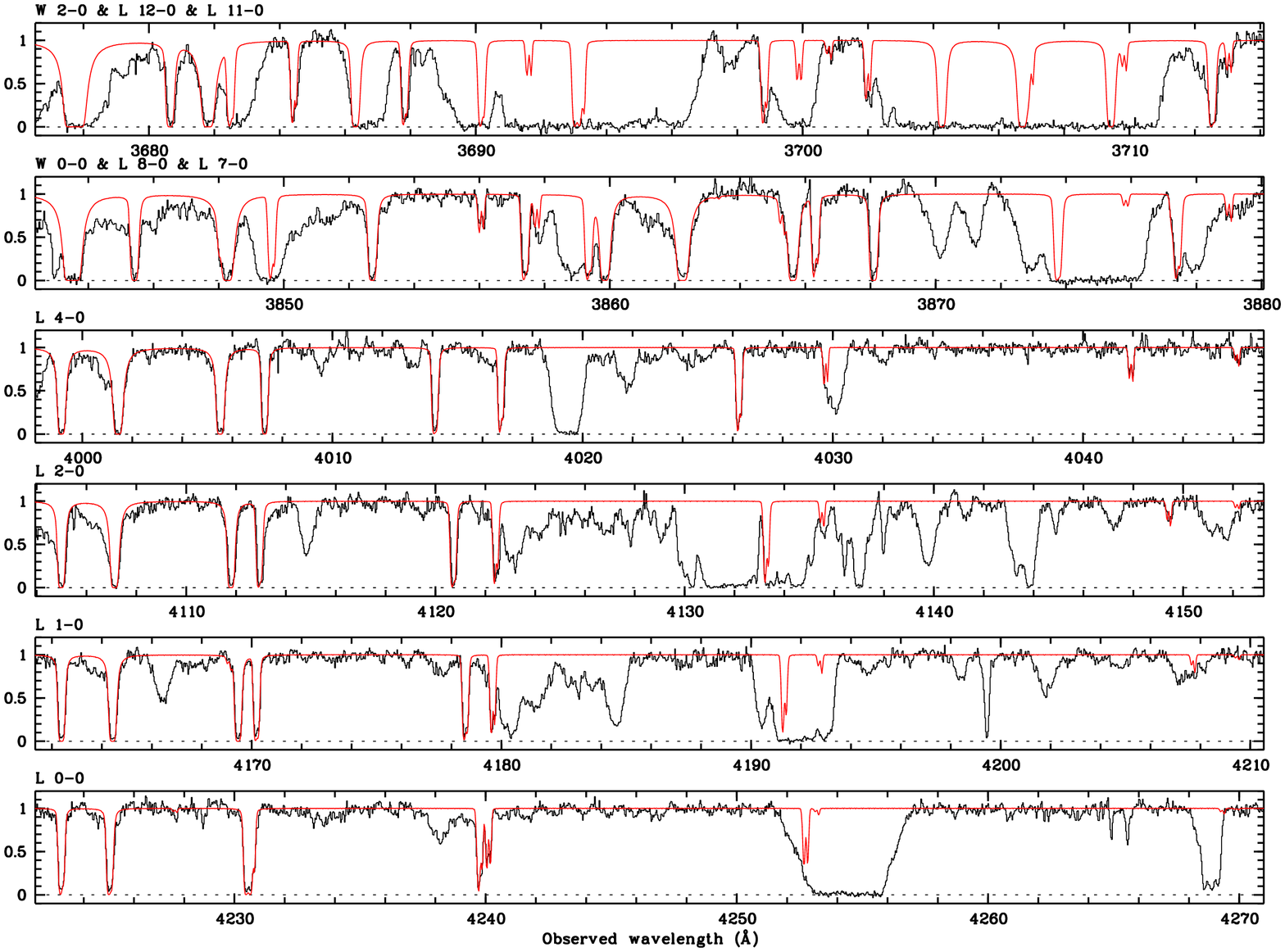,width=18.cm,height=10.cm,angle=270.}}}
\caption[]{Voigt profile fits to different rotational levels in 
the ground vibrational state of \h2 detected at \zabs = 2.5948 
toward Q\,0405$-$443 (top panels) and \zabs = 2.8111
toward Q\,0528$-250$ (bottom panels). 
The fit parameters are given in Table~\ref{h2tab}.
}
\label{h2fit}
\end{figure*}
\begin{table}
\caption{Results of Voigt profile fitting to \h2 absorption lines using the new data}
\begin{center}
\begin{tabular}{lcccc}
\hline
QSO & \zabs & J& $\log N$ (cm$^{-2}$) & $b$ (\kms)\\
\hline
\hline
Q\,$0405-443$  &2.59471&0 & 17.65$^{+0.07}_{-0.15}$ & 1.5$\pm$0.5\\
              &       &1 & 17.97$^{+0.05}_{-0.08}$ &            \\
              &       &2 & 15.93$^{+0.75}_{-0.42}$ &            \\
              &       &3 & 14.81$^{+0.14}_{-0.26}$ &            \\
              &       &4 & $\le13.55             $ &            \\
              &2.59486&0 & 15.13$^{+0.14}_{-0.10}$ & 1.1$\pm$0.1\\
              &       &1 & 15.24$^{+0.13}_{-0.02}$ &            \\
              &       &2 & 14.00$^{+0.01}_{-0.11}$ &            \\
              &       &3 & 13.93$^{+0.09}_{-0.05}$ &            \\
	      &       &4 & $\le13.55             $ &            \\
Q\,$0528-250$&2.81100&0 & 17.29$^{+0.08}_{-0.16}$ &2.8$\pm$0.5 \\
              &       &1 & 17.78$^{+0.10}_{-0.15}$ &            \\
              &       &2 & 16.74$^{+0.46}_{-0.46}$ &            \\
              &       &3 & 15.68$^{+0.44}_{-0.17}$ &            \\
              &       &4 & 14.13$^{+0.01}_{-0.01}$ &            \\
              &       &5 & 13.68$^{+0.01}_{-0.01}$ &            \\ 
              &2.81112&0 & 17.24$^{+0.05}_{-0.01}$ &1.0$\pm$0.3 \\
              &       &1 & 17.68$^{+0.07}_{-0.04}$ &            \\
              &       &2 & 16.98$^{+0.17}_{-0.36}$ &            \\
              &       &3 & 16.58$^{+0.15}_{-0.59}$ &            \\
              &       &4 & 14.32$^{+0.50}_{-0.15}$ &            \\
              &       &5 & 13.58$^{+0.02}_{-0.04}$ &            \\   
Q\,$1232+082$ &2.33772&0 & 19.34$^{+0.10}_{-0.10}$ &4.5$\pm$0.5\\
              &       &1 & 19.17$^{+0.10}_{-0.10}$ &            \\
              &       &2 & 16.70$^{+0.22}_{-0.22}$ &   \\
              &       &3 & 16.90$^{+0.32}_{-0.32}$ &\\
              &       &4 & 14.68$^{+0.05}_{-0.05}$ & \\
	      &       &5 & 14.45$^{+0.04}_{-0.04}$ & \\
\hline
\end{tabular}
\label{h2tab}
\end{center}
\end{table}
The  Ultraviolet and Visible Echelle Spectrograph
(UVES; Dekker et al. 2000), installed at the ESO VLT 8.2-m telescope unit
Kueyen on Mount Paranal in Chile was used to search for \h2 in a large
sample of DLAs. The sample and data reduction procedure are 
described in detail in Ledoux et al. (2003). Observations and details of Voigt
profile analysis of \h2 and metal line absorption lines toward Q\,0013$-$004, 
Q\,0551$-$366 and Q\,1232$+$082, along the lines of sight of which H$_2$ 
is detected, are described in, respectively, Petitjean et al. (2002), 
Ledoux et al. (2002) and Srianand et al. (2000). The Voigt profile fits 
to \h2 and other metal lines at \zabs = 3.024 toward Q\,$0347-383$, 
\zabs = 2.595 toward Q\,$0405-443$ and \zabs = 2.0868 toward Q\,$1444+014$ 
are discussed in Ledoux et al. (2003). For systems in which \h2 is not detected, 
Ledoux et al. (2003) have provided upper limits on $N$(\h2) together with mean
metallicities and depletion factors.

Recently, we have obtained additional higher spectral resolution
spectra ($R \sim$ 55,000) of Q\,0347$-$383 and Q\,$0405-443$
as a part of our ongoing programme on cosmic variation of
the electron-to-proton mass ratio (Petitjean et al. 2004).
Nine exposures of 1.5~h each were taken for each of the
quasars over six nights under sub-arcsec seeing conditions
in January 2002 and 2003 for Q\,0347$-$383 and Q\,$0405-443$,
respectively. We have also obtained additional data of
Q\,1232$+$082 to study the HD lines that are detected in
the DLA (Varshalovich et al. 2002). 
Spectra were reduced using the UVES pipeline  
and addition of individual exposures were performed using a sliding window  
and weighting the signal by the errors in each pixel. 
We detect a new \h2 component at \zabs = 2.59486 toward Q\,0405$-$443 
in addition to the strong component reported in Ledoux et al. (2003). 
We also present Voigt profile fits to
the \h2 lines in the \zabs = 2.811 system toward Q\,$0528-250$.
The single \h2 component seen in the lower spectral resolution 
CASPEC spectrum (Srianand \& Petitjean 1998) is resolved into two 
distinct components in our new UVES spectra. 
For both these systems the Voigt
profile fits to the \h2 Lyman and Werner band absorption
lines are shown
in Fig.~\ref{h2fit} and resulting parameters are summarised in
Table~\ref{h2tab}. This Table also gives the results of
Voigt profile fits to \h2 for the \zabs = 2.33772 toward Q\,1232$+$082.
\par
The main purpose of this paper is to provide a detail account of
C~{\sc i}, C~{\sc ii$^*$} and other metal lines in DLAs of our sample 
and extract physical conditions in conjunction with the \h2 content
reported in Ledoux et al. (2003). For the \zabs = 2.139
system toward Tol~1037$+$014 and the \zabs = 3.350 system toward 
Q\,$1117-1329$, we use the results presented in Srianand \& Petitjean (2001) and
P\'eroux et al. (2002) respectively. For the rest of the systems 
we give here the results of the multicomponent Voigt profile
fits.
For this we use a Voigt-profile fitting code that determines
the best fitting parameters (column density, velocity dispersion
and redshift) using $\chi^2$ minimization techniques (Chand et al. 2004).
We  use the oscillator strengths compiled in
Table~1 of Ledoux et al. (2003) for metal ions and those
given by Morton \& Dinerstein (1976) for H$_2$. 
In this article, we measure metallicities relative to
Solar, [X/H$]\equiv\log [N($X$)/N($H$)]-\log [N($X$)/N($H$)]_\odot$,
with either X$=$Zn, or S, or Si, and depletion factors of
iron, [X/Fe$]\equiv\log [N($X$)/N($Fe$)]-\log [N($X$)/N($Fe$)]_\odot$,
adopting the Solar abundances from Savage \& Sembach (1996).
\begin{table*}
{\caption{Excitation temperatures measured in individual \h2 components of DLAs in the 
sample of Ledoux et al. (2003)}}
\begin{tabular}{rcccccccc}
\hline
{QSO} &\zabs & log $N$(H~{\sc i})$^a$&log $N$(H$_2$) & $T_{01}$ (K) & $T_{02}$ (K) & $T_{13}$ (K) & OPR& 
$T$(OPR) (K)\\
\hline
\hline
Q $0013-004$ & 1.96685 & 20.83(0.05)&16.38$^{+0.03}_{-0.04}$& 300$^{+276}_{-96}$ & 766$^{+584}_{-244}$ &
             395$^{+59}_{-45} $ &1.85$\pm$0.28&114$^{+14}_{-12}$\\
             & 1.96822 && 16.54$^{+0.05}_{-0.05}$& 73$^{+7}_{-8}$     & 302$^{+33}_{-30}  $ &
             519$^{+59}_{-75} $ &0.64$\pm$0.09&66$^{+3}_{-4}$\\
Q $0347-383$ & 3.02489 &20.73(0.05)& 14.55$^{+0.09}_{-0.09}$&...                 & 740$^{+499}_{-212}$ &
             558$^{+100}_{-73}$ &3.45$\pm$0.47&$\ge 200$ \\
Q $0405-443$ & 2.59471&21.05(0.10) & 18.14$^{+0.07}_{-0.12}$&121$^{+10}_{-10}$   & 101$^{+24}_{-24}   $ &
             104$^{+4}_{-4}$  &2.10$\pm$0.30 & 127$^{+19}_{-15}$\\
             & 2.59486 & &15.51$^{+0.15}_{-0.07}$&91$^{+6}_{-6}$      & 118$^{+5}_{-5}     $ &
             219$^{+19}_{-19}$  &1.34$\pm$0.17 & 93$^{+6}_{-7}$\\
Q $0528-250$ & 2.81100 & 21.35(0.07)&17.93$^{+0.14}_{-0.20}$&167$^{+7}_{-7}$      & 190$^{+46}_{-46}$  &
             156$^{+11}_{-11}$  &2.47$\pm$0.36 &152$^{+50}_{-24}$\\
             & 2.81112 && 17.90$^{+0.11}_{-0.14}$&138$^{+12}_{-12}$   & 238$^{+46}_{-46}$   &
             278$^{+57}_{-57}$  & 1.83$\pm$0.33&$113^{+17}_{-13}$\\
Q $0551-366$ & 1.96168 & 20.70(0.08)&15.64$^{+0.40}_{-0.14}$&76$^{+7}_{-7}$      & 248$^{+52}_{-52}  $ &
             401$^{+73}_{-73}$  &0.90$\pm$0.16& 76$^{+8}_{-6}$\\
             & 1.96214 && 17.40$^{+0.65}_{-0.93}$&175$^{+88}_{-88}$    & 326$^{+65}_{-65}$ &
             446$^{+185}_{-185}$  &1.74$\pm$0.72&108$^{+42}_{-28}$\\
             & 1.96221 && 15.58$^{+0.03}_{-0.12}$&154$^{+24}_{-24}$    & 415$^{+25}_{-25}$   &
             593$^{+39}_{-39}$&1.86$\pm$0.23&115$^{+13}_{-14}$ \\
Q $1232+082$ & 2.33772 & 20.90(0.08)&19.57$^{+0.12}_{-0.12}$&67$^{+12}_{-12}$    & 67$^{+6}_{-6}$      &
             148$^{+22}_{-22}$  &$0.73\pm0.32$ &$71^{+12}_{-16}$\\
Q $1444+014$ & 2.08680 & 20.25(0.07)&16.49$^{+0.28}_{-0.11}$&285$^{+35}_{-35}$   & 205$^{+15}_{-15}$   &
             196$^{+14}_{-14}$  &3.59$\pm$0.42&$\ge 200$\\
             & 2.08696 & &18.15$^{+0.15}_{-0.15}$&193$^{+8}_{-8}$     & 148$^{+37}_{-37}$   &
             120$^{+1}_{-1}$    &3.20$\pm$0.28&$\ge 200$\\
\hline
\multicolumn{9}{l}{Note: The \zabs = 1.97296 and 1.97380 components toward Q\,0013$-$004 are not considered  here as the \h2 column densities }\\
\multicolumn{9}{l}{of the J~=~0 and J~=~1 levels are 
not known accurately due to saturation effects (see Petitjean et al. 2002)}\\
\multicolumn{9}{l}{$^a$ Integrated \hi column density in all the components 
(with and without \h2).
} 
\end{tabular}
\label{tab1}
\end{table*}
\section{Determination of physical parameters using H$_2$ level population}
%
In this section, we estimate different physical parameters 
from the column densities of \h2 in different J rotational levels.
\subsection{Kinetic temperature of the gas}
\begin{figure}
\flushleft{\vbox{
\psfig{figure=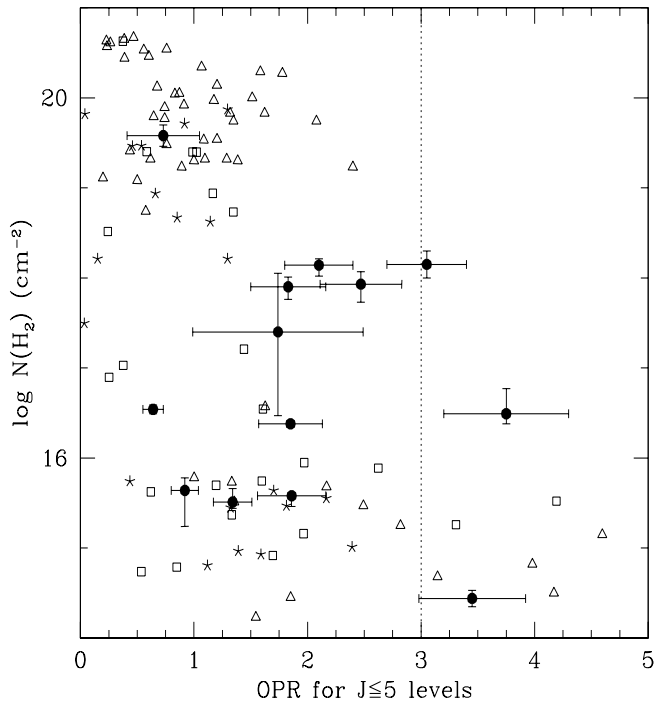,width=9.cm,height=10.cm,angle=0.}}}
\caption[]{
The Ortho-to-Para ratio (OPR) obtained using the populations of the
J$\leq 5$ rotational levels in individual components
of DLAs (black dots) is plotted against the total \h2 column density. 
Other data points are from Savage et al. (1977), Spitzer, Cochran \& 
Hirshfelf (1974) for the Galactic ISM (triangles), and 
Tumlinson et al. (2002) for the LMC (squares) and SMC (asterisks). 
The vertical short-dashed line shows 
the high temperature LTE limit of the OPR (i.e., OPR~=~3).
}
\label{fig2}
\end{figure}
\begin{figure}
\flushleft{\vbox{
\psfig{figure=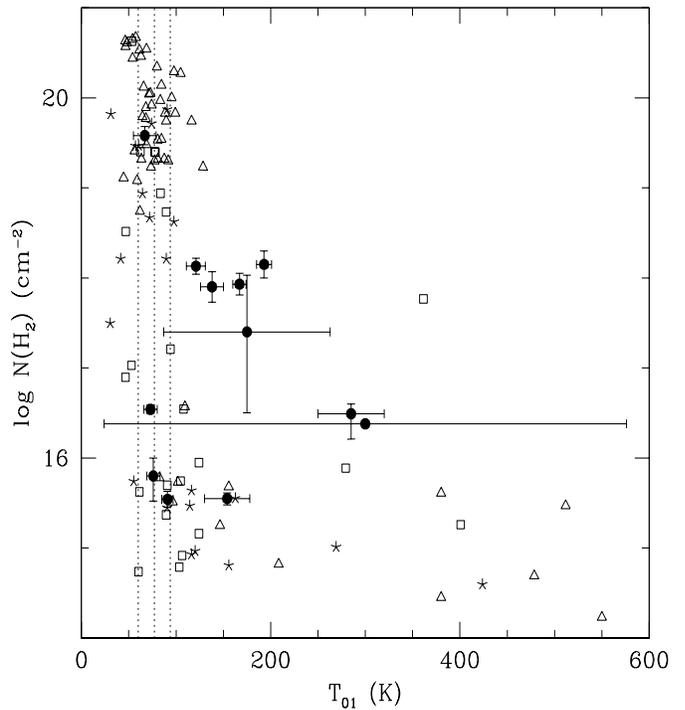,width=9.cm,height=10.cm,angle=0.}}}
\caption[]{
The rotational temperature $T_{01}$ obtained from the $N$(J=1)/$N$(J=0) 
ratio in DLAs (black dots) is plotted against the
total column density of molecular hydrogen. 
Other data points are from Savage et al. (1977), Spitzer, Cochran \& 
Hirshfelf (1974) for the Galactic ISM (triangles), and 
Tumlinson et al. (2002) for the LMC (squares) and SMC (asterisks). 
The vertical short-dashed lines show the mean and 1$\sigma$ range of 
$T_{01}$ measured by Savage et al. (1977) in the Galactic ISM.
}
\label{figext}
\end{figure}
It is a standard procedure, in ISM studies, to use the ortho-to-para
ratio (OPR) to infer the kinetic temperature of the gas assuming
local thermodynamic equilibrium, LTE 
(Tumlinson et al. 2002, and references there in). 
Indeed, recent numerical investigations suggest that the OPR is
a good tracer of the kinetic temperature over large regions of the
parameter space (Shaw et al. 2004). For completeness, we
first review our understanding of the OPR and outline the method for deriving
the kinetic temperature before applying the method to the data.
\subsubsection{General outline}
As the interconversion between para and ortho states involves a spin flip, 
it is not allowed for processes involving an isolated molecule (i.e., 
radiative processes cannot induce interconversion).
Ortho/para interconversion is only possible through (i) spin 
exchange induced by collisions with protons (with a rate coefficient   
in the range ${10^{-10}-10^{-9}~{\rm cm}^3 {\rm s^{-1}}}$; see Dalgarno, 
Black \& Weisheit 1973; Flower \& Watt 1984 and Gerlich 1990) or with 
hydrogen atoms (with a rate coefficient an order of magnitude less 
than that of protons; Mandy \& Martin 1993; Tin\'e et al. 1997)
and (ii) reactions on the surface of dust grains (Le Bourlot 2000).
In the case of local thermodynamic equilibrium (LTE),
\vskip 0.1 in
\begin{equation}
{\rm
OPR_{\rm LTE} = 3 {\sum_{J=odd} (2J+1)~\exp~[-BJ(J+1)/T]\over 
\sum_{J=even}(2J+1)~\exp~[-BJ(J+1)/T]}
}
\label{eqopr}
\end{equation} 
\vskip 0.1 in
\noindent
where, J, is the rotational quantum number, $B$ is the rotational constant
of \h2 ($B$ = 85.3 K), and ${T}$ is either the kinetic temperature of the 
gas (when OPR is governed by spin-exchange collisions) or the
formation temperature (when OPR is governed by H$_2$ formation
on the surface of dust grains with LTE distribution characterized by 
the formation temperature $T_{\rm form}$; see Sternberg \& Neufeld 1999;
Takahashi 2001). {The equilibrium temperature, T(OPR), can be obtained 
 using the observed value of OPR and Eq.~\ref{eqopr}. This will trace
the kinetic temperature of the gas if spin-exchange collisions 
are mainly responsible for the observed OPR.}
\par
If the gas is dense and cold and if most of the \h2~molecules are in 
the J~=~0 and J~=~1 levels then,
\vskip 0.1 in
\begin{equation}
{\rm
OPR_{\rm LTE} \sim {N(J=1)\over N(J=0)} = 9\times\exp (-170.5/T_{01}).
}
\label{t01}
\end{equation}
\vskip 0.1 in
\noindent
In the case of very optically thick molecular gas for which 
there is enough 
self-shielding, 
the $N$(J=1)/$N$(J=0) ratio can be maintained at its Boltzmann 
value and the excitation temperature, $T_{01}$, equals the
kinetic temperature. Savage et al. (1977)
measured a mean excitation temperature, T$_{01}$ = 77$\pm$17 K, for the galactic
ISM. This is consistent with the mean temperature of the ISM measured using 21 cm 
absorption lines. Thus it is widely believed that when there is sufficient 
shielding (i.e log $N$(H$_2$) cm$^{-2}$ $\ge$ 16.5), $T_{01}$ is a 
reasonably good tracer of the kinetic temperature. {This 
is because in the shielded region, \h2 photodissociation time-scale 
can be larger than the time-scale for charge exchange 
collision (Flower \& Watt, 1984). Also a recent
multi-wavelength study of Galactic sightlines show the 
T$_{01}$ measured in optically think cases closely follow the
spin temperature measured from 21 cm observations (see Roy et al 2005).}
\par
{The excitation temperature, $T_{ij}$, between different rotational levels
(say J= i and j) of a given species (either ortho or para \h2) can be obtained using,
\begin{equation}
{N(J=j)\over N(J=i)} = {2j+1\over 2i+1} \exp (-B[j(j+1)-i(i+1)]/T_{ij}). 
\label{text}
\end{equation}
Unlike OPR, this ratio can be altered by radiation pumping and formation
pumping in addition to collisions.
If collisions dominate the rotational excitation then $T_{ij}$ 
will be equal to T(OPR). Presence of formation
pumping and/or UV pumping will make $T_{ij}> T(OPR)$. In the following
section we discuss various temperature estimates from the DLA sample.
}
\subsubsection{Kinetic temperature of the \h2 components} 
In our sample, \h2 is only detected in J$\le5$ levels of the 
vibrational ground state. Thus we compute the OPR by summing the 
\h2 column densities for levels with J~$\le$~5. The observed value of 
the OPR for each DLA is given in column \#8 of Table~\ref{tab1}.
{
We calculate T(OPR) from the measured
OPR for individual systems using Eq.~\ref{eqopr}(see column \#9 of Table~\ref{tab1}).}
When the kinetic temperature (or formation temperature) is high
(i.e., $T\ge  200$ K) the OPR reaches 3, the value expected 
based on spin statistics. For a kinetic temperature similar to that seen
in the cold neutral medium of our Galaxy ($\simeq$ 80~K) 
the expected OPR under LTE assumption is less than 1. 
{From Table~\ref{tab1}, it is clear that the LTE temperatures 
measured from the OPR for DLAs are most of the times higher 
than 80 K (the mean found in the Galactic ISM.)}
\par
In Fig.~\ref{fig2} we plot the observed values of the OPR against the total 
\h2 column density  in the ISM (triangles), LMC (squares),
SMC (asterisks) and DLAs (circles with error-bars).
It is apparent that most of the OPR values in DLAs are 
significantly different from 
3 (see also column \#8 of Table \ref{tab1}). 
{ The distribution of the OPR as a function of $N$(H$_2$) in DLAs 
is consistent with that 
observed along Galaxy, LMC and SMC sightlines
(see Fig.~\ref{fig2}) when $\log$~N(H$_2$)$\le$16
and for rest of the components OPR in DLAs are systematically 
higher than that measured in the Galaxy, LMC and SMC. 
}
For example, OPR~$\ge$~3 is seen only 
along sightlines with low \h2~ optical depth (i.e $N$(\h2)~$\le10^{16}~
{\rm cm}^{-2}$) in the Galaxy, LMC and SMC (see Fig.~\ref{fig2}). 
On the contrary, out of the three DLA components with OPR~$\ge3$,
two, at \zabs = 2.08680 and \zabs = 2.08692 toward Q\,1444+014, are
optically thick in the Lyman band absorption lines.
\par
We next investigate the dependence of $T_{01}$ 
({measured using Eq.~\ref{t01}}) on the total H$_2$ column 
density (see Fig~\ref{figext}). Individual values measured in DLAs 
are listed in Table~\ref{tab1} (see column \#5). The large errors on both
$N$(H$_2$) and $T_{01}$ are mostly a consequence of the difficulty 
to measure the Doppler parameter when the lines are saturated.
In the case of the \zabs = 1.96685 component toward Q\,$0013-004$ the 
uncertainty is a consequence of line blending (see Petitjean et al. 2002).
The vertical dotted lines show the mean and 1$\sigma$ range of 
$T_{01}$ measured by Savage et al. (1977). The data points from the 
Magellanic clouds (Tumlinson et al. 2002) are consistent with this 
range (mean $T_{01}~=~82\pm21$ K). {As in the case of the
OPR, most of the measurements from DLAs with optically thick 
\h2 (i.e log N(\h2)$\ge16.5$) 
are well separated from that of the ISM and Magellanic 
clouds (Fig.~\ref{figext}) and the spread seen in the optically
thin case is consistent with that seen in local ISM.
}
Note that the system 
with lowest molecular content (\zabs = 3.02489
toward $0347-383$) has $N$(J=0) an order of magnitude lower than 
$N$(J =1). $T_{01}$ can not be computed in this case as the maximum expected 
column density ratio, $N$(J=1)/$N$(J=2), is 9 under LTE conditions
(see Eq.~\ref{t01}).
For the high optical depth clouds (i.e, log $N$(H$_2$) 
cm$^{-2}$ $\ge$ 16.5) in DLAs 
the mean $T_{01}$ is 153$\pm$78 K.
In most of the components the two temperatures $T_{01}$ and $T$(OPR) are consistent 
within errors. This is mainly because most of the \h2 molecules reside in the ground 
states. 
\par
{ In summary, if we assume LTE then $T$(OPR) and $T_{01}$ measured in DLAs (with
log~N(H$_2$)$\ge$16.5) at high redshift are on an average higher than that 
measured in the ISM, LMC and SMC sightlines. In this high N(\h2) range
$T_{01}$ is expected to trace the kinetic temperature. 
However, in the case of optically thin systems
the $T$(OPR) (or $T_{01}$) measured in DLAs are consistent with
that measured in LMC, SMC and Galactic sightlines.
%
Under the LTE assumption we find that \h2 components in DLAs have
kinetic temperatures in the range 100$-$200 K.
}
\subsection{Rotational excitation}
\begin{figure}
\flushleft{\vbox{
\psfig{figure=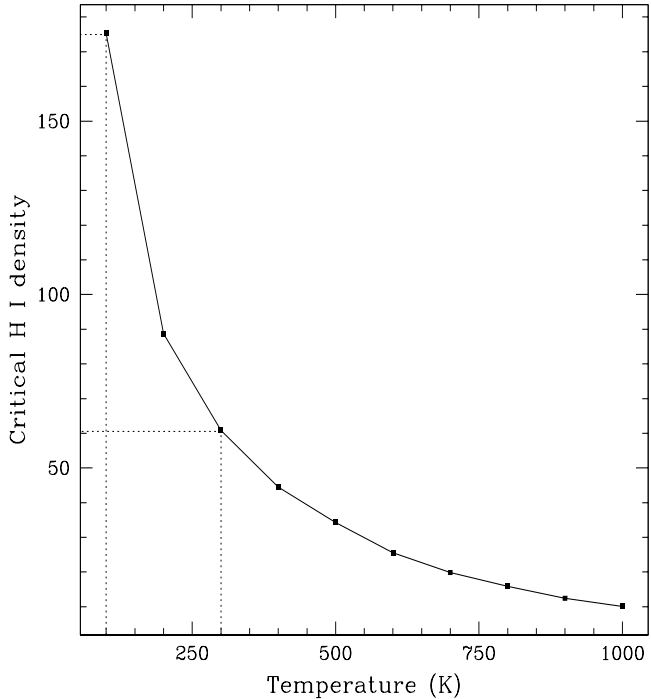,width=9.cm,height=10.cm,angle=0.}}}
\caption[]{Critical hydrogen density (${\rm n_H}$) as a function of kinetic
temperature for the thermalization of the J~=~0 and 2 levels. 
The curve is obtained by equating the collisional de-excitation 
and spontaneous decay rates. We have
used the collisional rates at low temperature given by 
Forrey et al. (1997). The spontaneous decay rates are from
Turner et al. (1977). It is clear from the short-dashed
lines that for 100$\le$T$\le$300 K the critical density is
175$\ge$\nh(cm$^{-3}$)$\ge$60.
}
\label{h2col}
\end{figure}
\begin{figure}
\flushleft{\vbox{
\psfig{figure=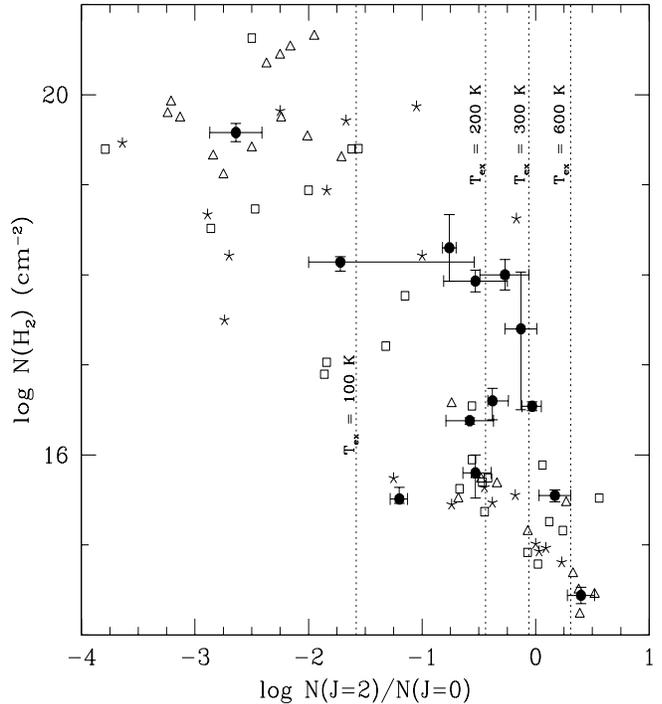,width=9.cm,height=10.cm,angle=0.}}}
\caption[]{The $N$(J=2)/$N$(J=0) ratio observed in individual DLA components
(black dots) is plotted 
against the total molecular hydrogen column density. 
Other data points are from Savage et al. (1977), Spitzer, Cochran \& 
Hirshfelf (1974) for the Galactic ISM (triangles), and 
Tumlinson et al. (2002) for the LMC (squares) and SMC (asterisks). 
The vertical dotted lines are the expected values 
of the column density ratio assuming the LTE for four different 
excitation temperatures. 
}
\label{fig3}
\end{figure}
The rotational level populations 
are affected by 
particle collisions, UV pumping, and formation pumping. While the
collisional excitation plays a significant role in populating the
low-J levels, those with J$\ge$3 are usually populated 
by formation processes and UV pumping. In what follows, we discuss the 
excitation of \h2 as seen in DLAs and compare with ISM, LMC and
SMC sightlines.
\subsubsection{Low-J excitation}  
The collisional contribution to the excitation of H$_2$ can be investigated 
by studying the $N$(J=2)/$N$(J=0) and  $N$(J=3)/$N$(J=1) ratios.
In general J = 2 and J = 3 levels can also be populated by deexcitation
of \h2~ formed in the high-J states (usually referred to as formation 
pumping) or through UV pumping. The collisional excitation rate for the 
J=0$\rightarrow$2 transition is about
an order of magnitude higher than that of the J=1$\rightarrow$3 transition for
kinetic temperatures in the range 100 to 300 K (Forrey et al. 1997).  
The spontaneous decay rate from J = 3 is an order of magnitude smaller that from
J = 2. This means that the ground and first excited states of para-\h2~ can 
be thermalised at lower densities compared to that of ortho-\h2. 
In Fig.~\ref{h2col}, we plot as a function of temperature the critical hydrogen 
density for which the collisional deexcitation rate becomes equal to the spontaneous 
decay rate for the J=2$\rightarrow$J=0 transition.
It is clear from this figure that the hydrogen density has to be high
(in the range 60$-$175 cm$^{-3}$) in order for the $N$(J=2)/$N$(J=0) ratio
to be equal to the LTE value corresponding to typical kinetic temperatures inferred
from the OPR (i.e 100 to 300 K). 
\begin{figure}
\flushleft{\vbox{
\psfig{figure=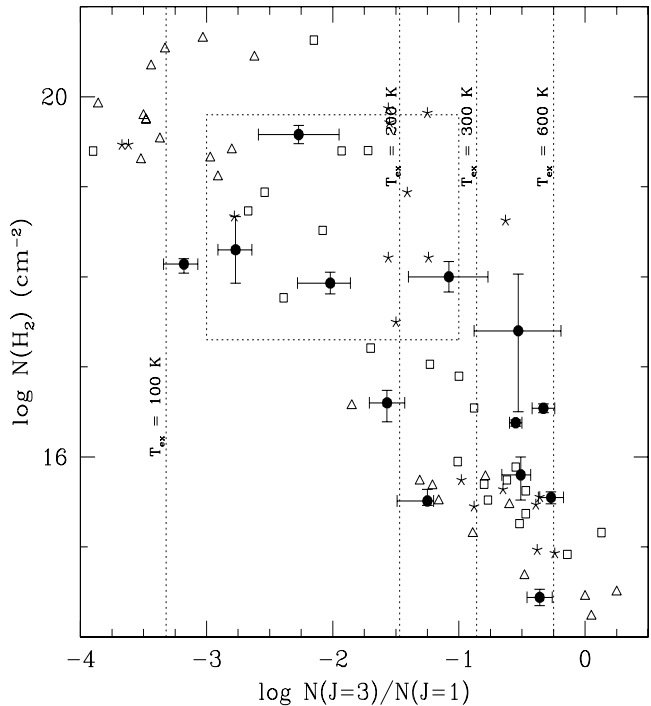,width=9.cm,height=10.cm,angle=0.}}}
\caption[]{The $N$(J=3)/$N$(J=1) ratio observed in individual DLA components
(black dots) is plotted 
against the total molecular hydrogen column density. The box drawn with
dotted lines gives the allowed range for the \zabs = 1.973 components 
toward Q\,0013$-$004.
Other data points are from Savage et al. (1977), Spitzer, Cochran \& 
Hirshfelf (1974) for the Galactic ISM (triangles), and 
Tumlinson et al. (2002) for the LMC (squares) and SMC (asterisks). 
The vertical dotted lines are the expected values of the column density 
ratio assuming the LTE for four different excitation temperatures.
}
\label{fig4}
\end{figure}
\par
In Fig.~\ref{fig3} we plot the ratio, $N$(J=2)/$N$(J=0), observed in DLAs, 
the Galaxy, LMC and SMC as a function of the total \h2 column 
density. The vertical dotted lines in the figure shows the expected values of
the ratio for four different excitation temperatures assuming LTE.
Values of the excitation temperature $T_{02}$ for individual DLA H$_2$ components
obtained using Eq.~\ref{text} are given in Table \ref{tab1}. 
The observed excitation temperatures are in the range 100 to 600 K with most of them 
at $T_{02}\simeq 150-300$ K.
If the level populations are in LTE then the required hydrogen
density to maintain the equilibrium is $65-150$ cm$^{-3}$
(see Fig.~\ref{h2col}). 
{
We can see from Fig.~\ref{fig3} 
that in DLAs where \h2~ is optically thick, the $N$(J=2)/$N$(J=0) ratio
is larger than that seen in similar gas of the Galactic ISM, LMC and
SMC. 
%
It can be seen from Table~\ref{tab1} that, in most of the DLAs, $T_{01}$ is lower than 
or equal to $T_{02}$ (see Table~\ref{tab1}). This is very much the case as well
in most of the sightlines through the ISM and Magellanic clouds. It is well 
known that, due to a lower value of the Einstein coefficient of the J = 2 
level compared to those of higher J levels, the UV and formation 
pumping processes can lead to enhancing the J = 2 level compared to the J = 0 
level. 
Thus the higher values of $N$(J=2)/$N$(J=0) seen
in DLAs can be explained by higher pressure in the gas and/or higher 
radiation field.}
\par
Fig.~\ref{fig4} gives the $N$(J=3)/$N$(J=1) ratio as a function of $N$(\h2).
The vertical dotted lines in the figure shows the expected value of the ratio 
for four different excitation temperatures under the LTE assumption.
The measurements in DLAs are consistent with local measurements and
the excitation temperature $T_{03}$ is in the range 100$-$680 K (see Table~\ref{tab1}).
%
\begin{figure}
\flushleft{\vbox{
\psfig{figure=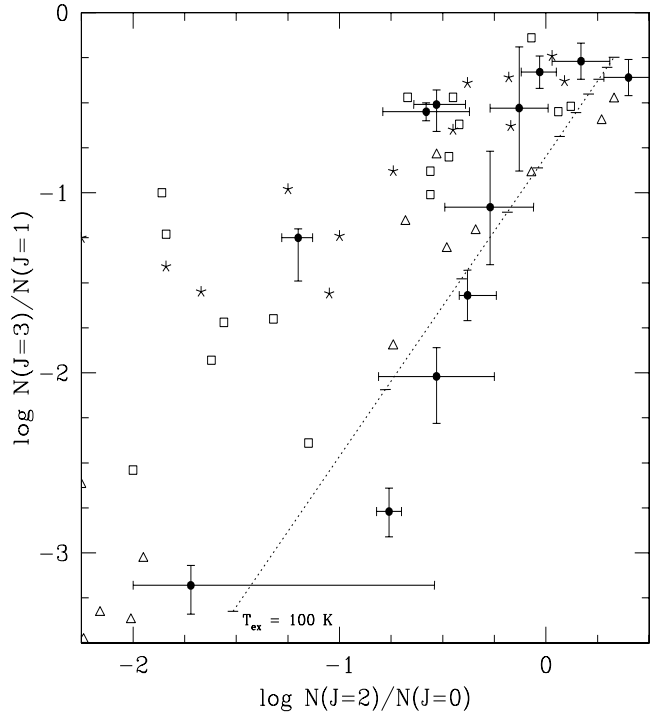,width=9.cm,height=10.cm,angle=0.}}}
\caption[]{The
$N$(J=3)/$N$(J=1) vs $N$(J=2)/$N$(J=0).
DLA measurements are indicated by black dots.
Other data points are from Savage et al. (1977), Spitzer, Cochran \& 
Hirshfelf (1974) for the Galactic ISM (triangles), and 
Tumlinson et al. (2002) for the LMC (squares) and SMC (asterisks).
The dotted line gives the expected relation under LTE with temperatures 
ranging from 100 to 600 K
(horizontal tick-marks show the values for different temperatures 
with 50 K steps).
}
\label{fig4a}
\end{figure}
In Fig.~\ref{fig4a} we plot 
the $N$(J=3)/$N$(J=1) ratio versus the $N$(J=2)/$N$(J=0) ratio. 
If formation and UV pumping contribute appreciably to populate the J~=~2 
and J~=~3 levels then we expect a tight relationship between the two quantities.
The dotted line in the figure gives the expected relationship between 
the ratios under LTE. In the case of sightlines through the Galactic ISM,
the LMC or SMC, the $N$(J=3)/$N$(J=1) ratio is higher than what is
expected from the $N$(J=2)/$N$(J=0) value under LTE (or, $T_{13}$ is higher than
$T_{02}$). In the case of DLAs, most of the components have $T_{13}$ 
close to $T_{02}$ (points are on top of the dotted line).  
Note that these excitation temperatures are different
from $T_{01}$. This clearly means that UV pumping and formation
pumping are not negligible even for the excitation of the low J levels.  
%
The nature of the local radiation field
can be probed using excitations of J$\ge$ 3 levels. 
This is what we do in the following Section.
\par
\begin{figure}
\flushleft{\vbox{
\psfig{figure=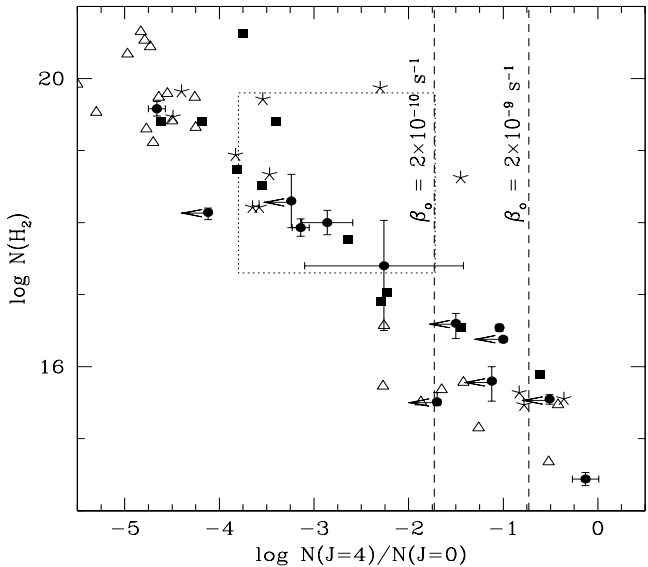,width=9.cm,height=8.cm,angle=0.}}}
\caption[]{The $N$(J=4)/$N$(J=0) ratio
is plotted against the total molecular hydrogen column density.
DLA measurements are indicated by black dots.
Other data points are from Savage et al. (1977), Spitzer, Cochran \& 
Hirshfelf (1974) for the Galactic ISM (triangles), and 
Tumlinson et al. (2002) for the LMC (squares) and SMC (asterisks).
The box with dotted lines gives the allowed range for the \zabs = 1.973 component
toward Q\,0013$-$004. The vertical dashed lines are the expected values of the 
ratio for two different values of the photo-absorption rate ($\beta_0$).
}
\label{fig6}
\end{figure}
\subsubsection{UV radiation field: High-J excitation}
It is known that in the photodissociation regions (PDRs) the J~=~4 and J~=~5 
rotational levels are populated predominantly by cascades following 
the formation of excited molecules and UV pumping from the low-J states. 
As radiative decay time-scales for these levels are very short compared
to the collisional time-scales, spontaneous decay is the main deexitation 
process. Among the two populating processes the UV pumping is an optical depth 
dependent process while the formation pumping is independent of optical depth.
In an optically thick cloud, UV pumping is efficient in a thin shell 
surrounding the cloud. In the interior of the cloud, UV pumping
becomes important only when the column density becomes very large
(i.e. through absorption in the damping wings).
\par
In Fig.~\ref{fig6} we plot log~$N$(\h2) as a function of log~$N$(J=4)/$N$(J=0)
as measured in DLAs and along the ISM, LMC and SMC sightlines.
As expected, a strong anti-correlation is present in the data, including DLAs.
\h2 absorption lines in DLA systems have no strong overlapping wings. Therefore the 
high-J excitation 
is mostly due to photo-absorption in the systems with log $N$(\h2) $\le$16.5 
and to \h2 formation in the systems with higher column densities. Following analytic 
prescription by Jura (1975) we can write,
\vskip 0.1in
\begin{equation}
{\tiny
p_{4,0}\beta(0)n(H_2,J=0)+0.24 Rn(H)n = A(4\rightarrow 2)n(H_2,J=4)}
\label{uvpump}
\end{equation}
\vskip 0.1in
\noindent
Here, $\beta(0)$, $p_{4,0}$ are, respectively, the photo-absorption rate in the Lyman and Werner 
bands and the pumping efficiency from J~=~0 to J~=~4; ${\rm A(4\rightarrow 2)}$
is the spontaneous transition probability between J~=~4 and J~=~2 and $R$ is the 
formation rate of \h2. Neglecting the second term in the left hand side of 
Eq.~\ref{uvpump} leads to a conservative upper limit on the UV radiation field.
The vertical dashed lines in Fig.~\ref{fig6} represent
the corresponding predicted values of the $N$(J=4)/$N$(J=1) ratio 
for $\beta(0)=2\times10^{-10}$s$^{-1}$ (that is, approximatively the mean radiation field 
in the ISM) and $\beta(0)=2\times10^{-9}$s$^{-1}$.
It can be seen from Fig.~\ref{fig6} that for log $N$(\h2) less than 16.5
the  $N$(J=4)/$N$(J=0) ratio in DLAs is of the order of or slightly
higher than that seen in  the ISM of our Galaxy. Quantitatively the
upper limits in most of the systems are consistent with 
$2\times10^{-10}\le\beta(0)\le2\times10^{-9}$s$^{-1}$. 
This probably means the optically thin \h2 components without
detectable \h2 absorption lines from the J~=~4 state arise in gas 
embedded in a
UV field with intensity similar to (or slightly higher than) that of 
the mean ISM field. 
\par
There are two optically thin components in our sample
(\zabs = 1.96822 toward Q\,0013$-$004 and 3.02489 toward Q\,0347$-$383)
that show detectable J~=~4 \h2 absorption lines. Detailed analysis of these
component suggests an ambient field intensity consistent with few times
the mean ISM field intensity (Petitjean et al. 2002; Levshakov et al. 2002).
The same conclusion was derived by Reimers et al. (2003) for the optically thin 
\h2 component at \zabs = 1.15 system toward HE~$0515-4414$.
\par
The above ratio has similar values at high log~$N$(\h2) in DLAs and 
in our Galaxy. This is a hint for the formation pumping in 
DLAs with high $N$(\h2) being similar to the local one. 
There are two optically thick components (at \zabs = 2.59471 toward
Q\,0405$-$443 and \zabs = 2.08696 toward
Q\,1444$+$014) that do not show detectable absorption lines 
from the J~=~4 state.
In the latter system the ratio $N$(J=4)/$N$(J=0)$\le 10^{-4}$. This is 
much lower than the values seen in the ISM at similar total $N$(\h2) and 
could be a consequence of lower \h2 formation rate in this system. 
High values of the radiation field intensity were inferred for some of the
optically thick components when the contribution of the second term in 
Eq.~\ref{uvpump} is estimated using the average metallicity and dust depletion
(Ge \& Bechtold 1997; Petitjean et al. 2000; Ge, Bechtold \& Kulkarni 2001).
\section {Analysis of Carbon absorption lines}           
As the ionization potential of C~{\sc i} is 11.2 eV, the ionization state of Carbon 
is sensitive to the same photons that destroy \h2. Therefore, 
C~{\sc i} is usually a good tracer of the physical conditions in the molecular gas 
(see however Srianand \& Petitjean 1998). In what follows we investigate 
the relationship between the detectability of C~{\sc i} absorption line
and other measurable 
quantities in our spectra. We derive additional constraints on the physical conditions 
in DLAs using C~{\sc i} fine-structure absorption lines.
\subsection{Detectability of C~{\sc i} absorption lines}
The results of simultaneous Voigt
profile fitting to C~{\sc i}, C~{\sc i}$^*$ and C~{\sc i}$^{**}$ absorption lines
in our sample are summarised in Table~\ref{tabphy}.
\par
In the interstellar medium of our Galaxy, all clouds with 
log~$N$(H~{\sc i})~$\ge 21$ 
have log~$N$(H$_2$)~$>19$ and log~$N$(C~{\sc i})~$>14$ (Jenkins \& Shaya 1979; 
Jenkins, Jura \& Loewenstein 1983). 
In our sample, C~{\sc i} absorption lines are detected in most of the DLAs that show \h2 
absorption lines(see also Ge \& Bechtold 1999). There are three exceptions: the components at 
\zabs = 2.59471 and 2.59486 toward Q\,0405$-$443 and at \zabs = 2.81100 toward 
PKS~0528$-$250. These C~{\sc i} non-detections are surprising as the \h2 absorption lines
from these components are optically thick so that C~{\sc i} is expected to be conspicuous. 
\par
Usually, DLAs in which no \h2 is detected through the whole profile 
do not show any detectable C~{\sc i} absorption line 
(with a typical upper limit 
of 10$^{12}$ cm$^{-2}$). The only exception is 
the high-metallicity sub-DLA at \zabs = 2.139 toward Tol~1037$-$270 
(see Srianand \& 
Petitjean 2001). On the contrary, in DLAs where \h2 is detected, some components
show detectable C~{\sc i} absorption line without detectable molecular absorption 
($N$(\h2)$\le10^{14}$cm$^{-2}$). This is the case in Q\,0013$-$004 (Petitjean 
et al. 2002) and Q\,0551$-$366 (Ledoux et al. 2002).
\par
Note that C~{\sc i} is also detected at \zabs = 2.28749 toward Q\,2332$-$094
but the presence of \h2 molecules can not be probed in this system 
due to the presence of an intervening Lyman limit system. 
The sub-DLA at $z$~=~1.15 toward HE~0515$-$4414 shows C~{\sc i} and
\h2 absorption lines (Quast et al. 2002; Reimers et al. 2003).
C~{\sc i} absorption lines have also been detected at \zabs = 1.776 system
toward Q\,1331+170 (Chaffee et al. 1988). {Presence of \h2 in this
system is recently reported (Cui et al., 2004)}.
%
\begin{figure}
\flushleft{\vbox{
\psfig{figure=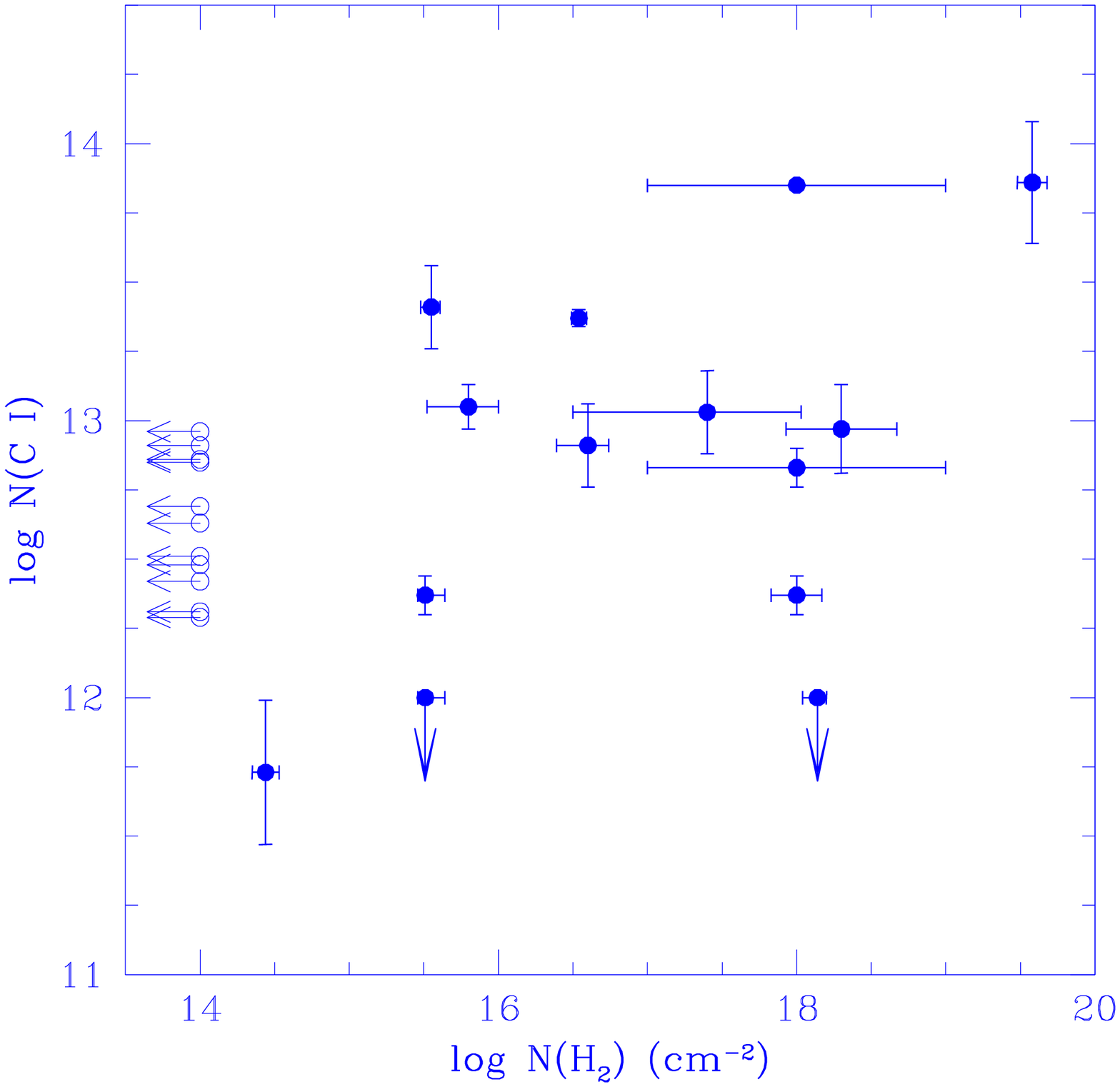,width=8.cm,height=8.cm,angle=0.}}}
\caption[]{The column density of C~{\sc i} measured in
individual components of DLAs is plotted as a function of $N$(\h2).
The filled circles with error-bars are points from the components that 
show detectable \h2 absorption lines. The open circles are the measurements from
components that show C~{\sc i} without detectable \h2. 
}
\label{c1fig1}
\end{figure}
\begin{table*}
\caption {DLA systems with detected C\,{\sc i} absorption lines in the 
sample of Ledoux et al. (2003)}
{\tiny
\begin{tabular}{cccccccccc} 
\hline
Quasar/ & \zabs &$\log N(H_2)$ &$\log N($C\,{\sc i}$)$ & $\log N($C\,{\sc i}\,$^\star)$ &
$\log N($C\,{\sc i}\,$^{\star\star })$ & $b$(km s$^{-1}$)  & T$_{\rm CMB}$(K) & n$_{\rm H}$(cm$^{-3}$) 
&P/k (cm$^{-3}$ K)\\
log N(H~{\sc i})\\
\hline
\hline
Q\,0013$-$004& 1.96679&$^a$16.38$^{+0.03}_{-0.04}$& $13.06\pm 0.04$&$12.72\pm 0.05$&....            &3.1$\pm$0.6 &$<13.8$&$20-70$&2000$-$7000\\
\multicolumn {1}{c}{20.83(0.05)}& 1.96691&....& $12.86\pm 0.03$&$12.55\pm 0.07$&....                &3.0$\pm$0.2&$<14.5$&$20-80$&2000$-$8000\\
            & 1.96706&....& $12.31\pm 0.07$&$12.26\pm 0.11$&....            &4.93$\pm$1.44 &....    &....&....\\ 
            & 1.96822 &$16.54^{+0.05}_{-0.05}$& $13.04\pm 0.01$&$13.10\pm 0.01$&$12.77\pm 0.02$     &4.3$\pm$0.2 &$<20.0$&$170-200$&$11730-12400$\\ 
            & 1.97280 &$^b$17.48$-$19.78& $13.08\pm 0.01$&$12.77\pm 0.03$&....                      &6.0$\pm$0.3 &$<13.0$&$40-60$&4000$-$6000\\ 
            & 1.97296 & &$13.44\pm 0.01$&$13.20\pm 0.01$&$12.53\pm 0.04$                            &6.2$\pm$0.2 &$<13.5$&$50-65$&$5000-6500$\\ 
            & 1.97316 && $12.66\pm 0.03$&$12.33\pm 0.10$&....                                       &7.0$\pm$0.8 &$<15.0$&$10-85$&1000$-$8500\\ 
            & 1.97365 &$^c$17.30$-$19.60& $12.32\pm 0.05$&$   < 12.11$&....                         &5.7$\pm$0.4 &$<15.0$&$<85$&$<8500$\\ 
            & 1.97382 &....& $12.40\pm 0.04$&$12.04\pm 0.13$&....                                   &4.5$\pm$0.4 &$<15.5$&$<95$ &$<9500$\\ 
            & 1.97399 &....& $11.91\pm 0.13$&....           &....                                   &7.3$\pm$1.0 &....     &....&....\\
            & 1.97417 &....& $11.49\pm 0.25$&....           &....                                   &0.4$\pm$0.3 &....     &....&....\\
            & 1.96737&$<14.0$ & $12.91\pm 0.03$&$12.77\pm 0.06$&....                               &14.4$\pm$1.1 &$<17.5$&$40-135$&4000$-$13500\\ 
            & 1.96763&$<14.0$ & $12.85\pm 0.05$&$12.68\pm 0.09$&....                               &27.1$\pm$3.4 &$<19.3$&$20-180$&2000$-$18000\\ 
            & 1.97109 &$<14.0$& $12.28\pm 0.03$&$< 12.11$   &....                                  &10.0$\pm$0.9 &$<21.0$&$<200$&$<20000$\\ 
            & 1.97144 &$<14.0$& $12.96\pm 0.03$&$12.91\pm 0.03$&....                               &13.4$\pm$0.9 &$<18.5$&$65-160$& 6500$-$16000\\ 
Q\,0347$-$383& 3.02485 &14.55$^{+0.09}_{-0.09}$& $11.73\pm 0.26$&$<11.50     $&$<11.75$            &$ 4.9\pm0.3$   &$<17.0$&$< 48$&$<4800$\\ 
\multicolumn {1}{c}{20.73(0.05)}\\
Q\,0405$-$443& 2.59474 &18.14$^{+0.07}_{-0.12}$& $<12.23     $& ....          & ....          &....           &....    &.... &.... \\ 
\multicolumn {1}{c}{21.05(0.10)}& 2.59485 &15.51$^{+0.15}_{-0.07}$& $<11.90     $& ....          & ....          &....           &....    & .... &....  \\ 
Q\,0528$-$250& 2.81100 &17.93$^{+0.14}_{-0.20}$& $< 12.00$&....          & ....          &....           &....    & .... &....  \\ 
\multicolumn {1}{c}{21.35(0.10)} & 2.81112 &17.90$^{+0.11}_{-0.14}$& $12.36\pm0.10$ & $12.30\pm0.10$ &..... & 0.6$\pm$0.1 &$<17.0$&$25-270$&$3250-17000$\\
Q\,0551$-$366& 1.96152 &$<14.0$& $12.69\pm 0.07$&$<12.18$    &$<11.94$     &$ 4.3\pm1.4$   &$<10.5$&$<17$ &$<1700$\\ 
\multicolumn {1}{c}{20.70(0.10)} & 1.96168 &15.80$^{+0.40}_{-0.14}$& $12.64\pm 0.07$&$12.84\pm 0.07$&$<12.16$     &$ 2.1\pm0.7$   &$<22.3$&$170-185$&$12950-14280$\\ 
            & 1.96180 & $<14.0$&$12.42\pm 0.13$&$<12.18$       &$<11.94$     &$ 3.9\pm2.3$   &$<12.9$&$<56 $& $<5600$\\ 
            & 1.96214 & $17.40^{+0.65}_{-0.93}$&$12.66\pm 0.12$&$12.69\pm 0.11$&$12.11\pm0.34$ &$ 2.1\pm0.8$   &$<18.7$&$55-390$&$8250-30400$\\ 
            & 1.96221 &$15.58^{+0.03}_{-0.12}$& $13.16\pm 0.06$&$12.98\pm 0.09$&$12.26\pm0.36$ &$12.8\pm1.7$   &$<14.5$&$30-150$&$3840-15150$\\ 
            & 1.96268 &$<14.0$& $12.63\pm 0.08$&$<12.18$       &$<11.94$       &$ 4.0\pm1.8$   &$<10.0$&$<25 $&$<2500$\\  
Q\,1037$-$270& 2.13900&$<14.0$&$12.51\pm0.02 $&$<12.48$    &....            &15.5$\pm$0.6  &$<16.4$&$<105$&$<10500$\\ 
\multicolumn {1}{c}{19.70(0.05)}& 2.13940&$<14.0$&$12.48\pm0.02$ &$<12.30$    &....            &4.5$\pm$1.2 &$<13.2$&$<65 $&$<6500$\\ 
Q\,1232$+$082& 2.33771 &19.57$^{+0.12}_{-0.12}$& $13.86\pm 0.22$&$13.43\pm 0.07$&$12.63\pm0.22$ &$1.7\pm0.1$   &$<14$&$40-60$&$2320-5280$\\
\multicolumn {1}{c}{20.90(0.08)}\\
Q\,1444$+$014& 2.08679 &16.60$^{+0.28}_{-0.11}$& $12.67\pm 0.09$&$12.52\pm 0.14$& ....          &$11.6\pm2.8$   &$<18.5$&$20-110$&$6560-27280$\\ 
\multicolumn {1}{c}{20.25(0.07)} & 2.08692 &18.15$^{+0.15}_{-0.15}$& $12.82\pm 0.11$&$12.42\pm 0.12$& ....          &$1.1\pm0.3$    &$<14.0$&$ 4-54 $&$824-10260$\\ 
Q\,2332$-$094& 2.28749 &....& $13.33\pm 0.03$&$13.14\pm 0.03$&$12.34\pm0.11$ &$3.8\pm0.3$    &$<14.6$&$45-75 $&$4500-7500$\\ 
\multicolumn {1}{c}{20.25(0.07)}\\
\hline
\multicolumn {10}{l}{
$^a$ total \h2 column density in three components at \zabs = 1.96679, 1.96691 and 1.96706 is quoted.}\\
\multicolumn {10}{l}{$^b$ total \h2 column density in three components
at \zabs = 1.97280, 1.97296 and 1.97316.}\\
\multicolumn {10}{l}{$^c$ total \h2 column density in four 
components at \zabs = 1.97365, 1.97382, 1.97399 and 1.97417 is quoted.}\\ 
\end{tabular}
}
\label{tabphy}
\end{table*}
\begin{figure}
\flushleft{\vbox{
\psfig{figure=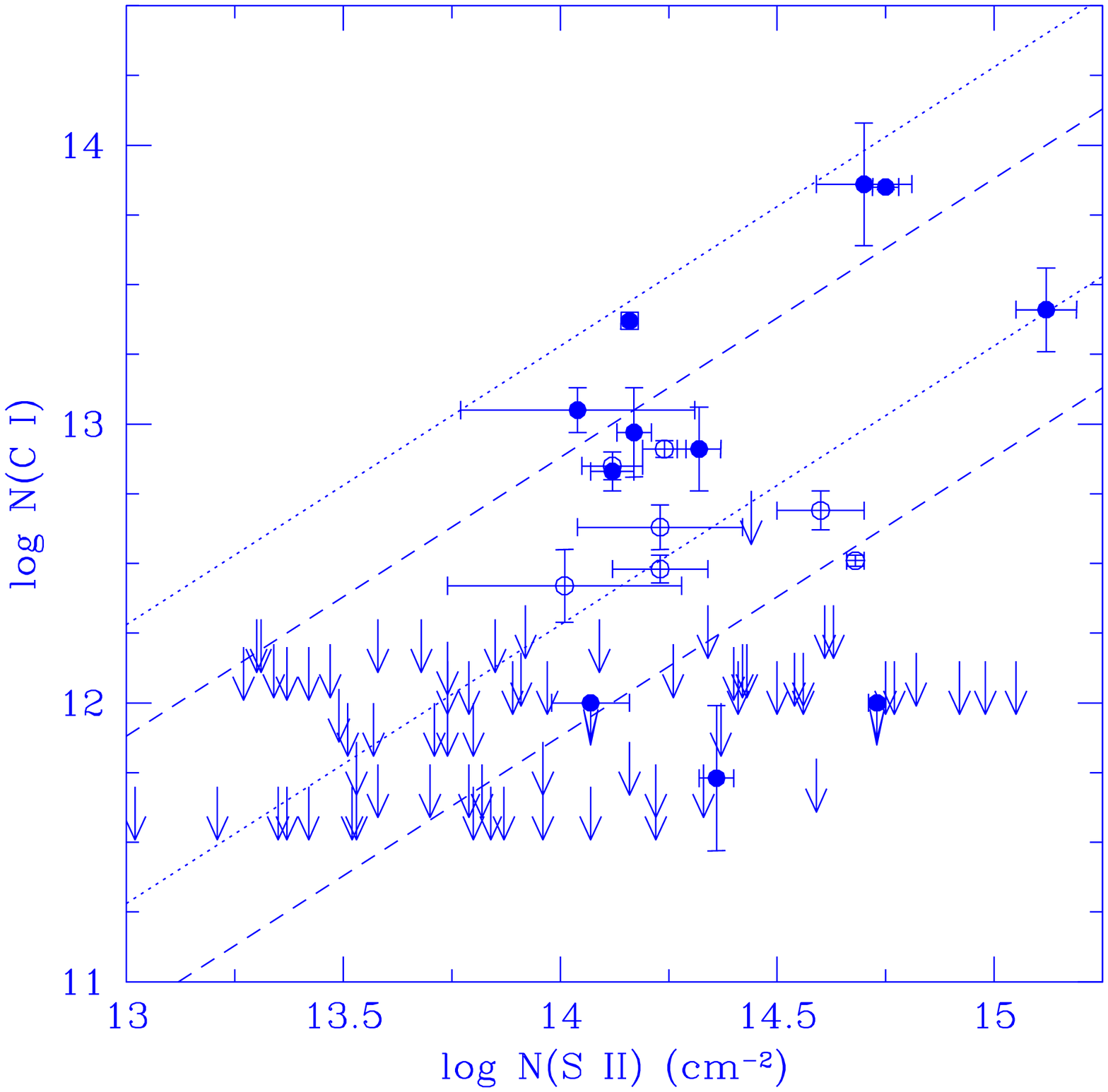,width=8.cm,height=8.cm,angle=0.}}}
\caption[]{The C~{\sc i} column density measured in individual components of
DLAs is plotted as a function of the S~{\sc ii} column density. The filled circles 
with error-bars are points from components that show detectable 
\h2 absorption lines. 
The open circles are the measurements from components that show C~{\sc i} but no 
detectable \h2.  The arrows are the upper limits from systems where there is no \h2 and/or 
C~{\sc i} absorption lines. The two dotted lines give the expected correlation for 
log $N$(C~{\sc i})/$N$(C~{\sc ii}) = $-3$ (lower line) and $-2$ (upper line) 
respectively, when solar relative abundances are used and it is assumed that there
is no depletion of Carbon relative to Sulfur. The short-dashed lines give the 
same correlations when a depletion of Carbon relative to Sulfur of 0.4 dex is further
assumed. The C~{\sc i} detections are consistent with what one expects in the case of the CNM. 
}
\label{c1fig2}
\end{figure}
\subsubsection{Dependence on \h2 column density}
Jenkins \& Shaya (1979) found $N$(C~{\sc i}) does not scale linearly with either 
of $N$(H~{\sc i}), $N$(H$_2$) or $N$(H$_{\rm total}$) in the Galactic ISM. 
They explained this behavior as a result of strong differences in the response 
of C~{\sc i}, H~{\sc i} and H$_2$ to physical conditions (electron density, 
temperature etc...), coupled with marked variations of these conditions from 
one cloud to the other. In Fig.~\ref{c1fig1}, we plot the C~{\sc i} column 
density as a function of \h2 column density in individual components. 
Among the systems that show \h2 absorption lines(filled circles with error-bars) there 
is no clear trend between $N$(\h2) and $N$(C~{\sc i}) even though the presence 
of C~{\sc i} absorption lines usually indicate the presence of \h2 (see discussion above). 
\par
\subsubsection{The Carbon ionization state}
The probability of detecting C~{\sc i} is expected to be higher
in systems with higher $N$(H~{\sc i}) and/or metallicity. 
Ideally, we would like therefore to know $N$(H~{\sc i}) for each individual C~{\sc i} 
components. This is not possible as all components are definitely blended in one strong 
H~{\sc i} DLA absorption line. {Estimation of N(H~{\sc i}) is possible when the 
\h2 component is well separated from the rest of the components (as in
\zabs = 1.96822 toward Q\,0013$-$004) or when 21 cm observations are
available (as in the case of \h2 components toward Q\,0528$-$250).
Note that the presence of very strong C~{\sc i} absorption line in 
the component at \zabs = 1.96822 
toward Q\,0013$-$004 (that has log~N(H~{\sc i})$\le$19.4) is mainly
due to high metallicity (Petitjean et al. 2002). Whereas the absence of 
C~{\sc i} in the component at \zabs = 2.81100 and the 
weakness of C~{\sc i} line of the component at
\zabs = 2.81112 component toward Q\,0528$-$250 are probably due to
excess radiation field from the QSO.} 

\par
The ionization state of Carbon is difficult to determine as the
C~{\sc ii}$\lambda1334$ absorption line is usually highly saturated.
We can however partly overcome this difficulty assuming that conditions 
are fairly homogeneous in the DLA system. 
Under the assumptions that the enrichment
of Carbon follows that of $\alpha-$elements and that the relative depletion between
Sulfur and Carbon is negligible, we can use the well determined $N$(S~{\sc ii}) as an 
indicator of $N$~(C~{\sc ii}) (see Fig.~\ref{c1fig2}). In the sun, the Carbon abundance 
is 1.28 dex higher than that of Sulfur and typical depletion of Carbon relative to Sulfur 
in the Cold ISM is 0.4 dex (see Table 5 of Welty et al. 1999). The two dotted lines in 
Fig.~\ref{c1fig2} give the expected correlation for log $N$(C~{\sc i})/$N$(C~{\sc ii}) = $-3$
(lower line) and $-2$ (upper line) respectively, when relative solar abundances are used
and it is assumed that there is no depletion of Carbon relative to Sulfur. The short-dashed 
lines give the same correlations when a depletion of Carbon relative to Sulfur of 0.4 dex is
further assumed. If the absorbing gas originates from the CNM then 
log $N$(C~{\sc i})/$N$(C~{\sc ii}) is expected to be more than $-3$ (see Fig.~3 of 
Liszt 2002). Therefore, within uncertainties due to depletion, it is apparent 
from  Fig.~\ref{c1fig2} that the DLA components with C~{\sc i} detections have a ionization 
state consistent with them originating from the CNM. 
\par
It is interesting to note that the distribution of $N$(S~{\sc ii}) is somewhat 
similar for components with both \h2 and C~{\sc i} absorption lines (filled circles), 
and for components with C~{\sc i} but no \h2 absorption lines (open circles). 
However C~{\sc i} column densities are typically lower in the components without \h2
suggesting, as expected, that when \h2 is seen, the C~{\sc i}/C~{\sc ii} ratio is larger.
\par
Most of the upper limits on C~{\sc i} 
are consistent with $N$(C~{\sc i})/$N$(C~{\sc ii})~$\le$~$-3$ (see Fig.~\ref{c1fig2}). 
This can mean either that the relative depletion of Carbon compared to Sulfur 
is larger than 0.4 dex in the CNM, which is unlikely, or that most of the DLA systems
originate from the warm neutral medium (WNM) or warm ionized 
medium (WIM) where the above ratio can be as low as $10^{-4}$. 
\begin{figure}
\psfig{figure=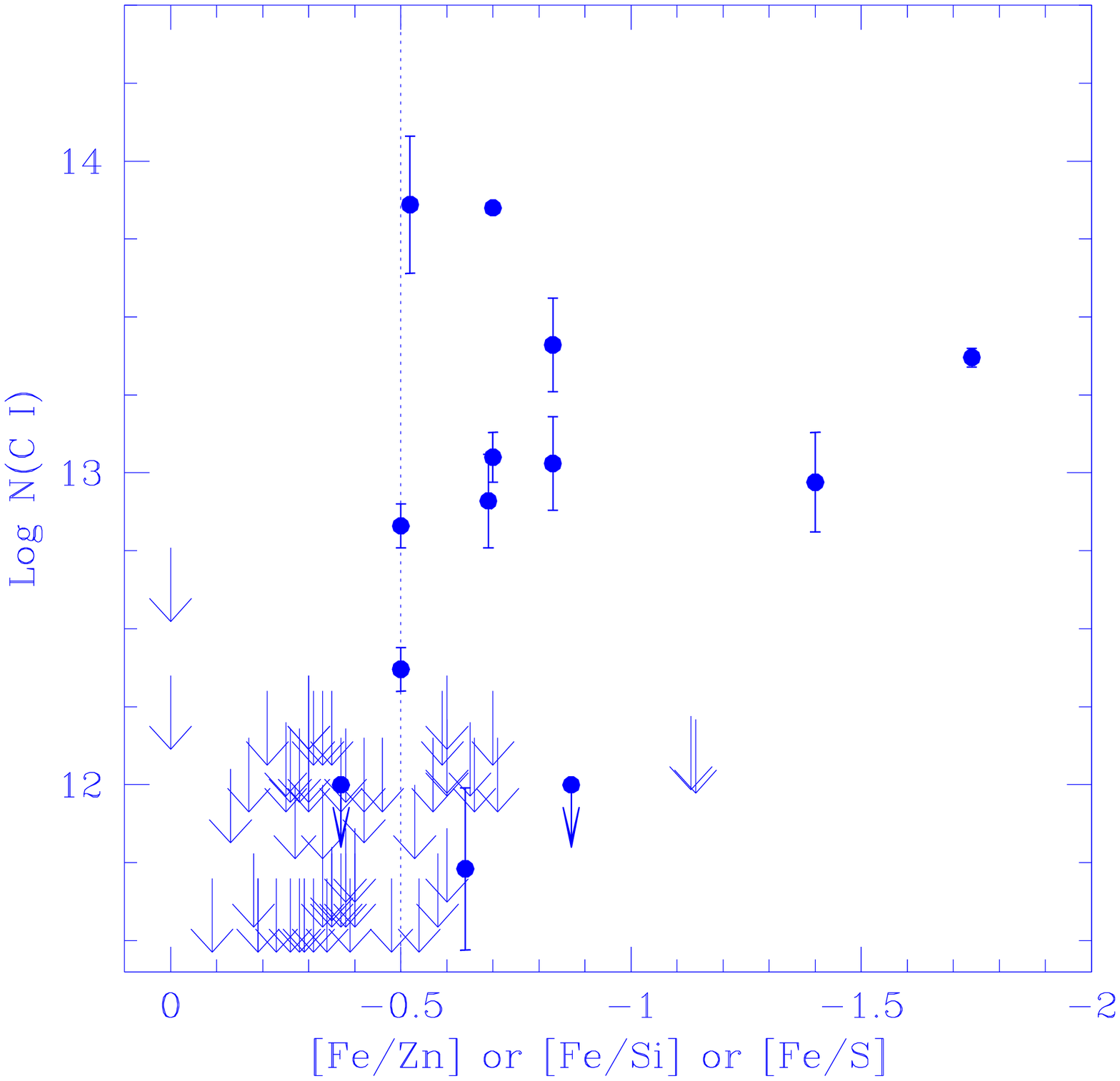,width=8.cm,height=8.cm,angle=0.}
\caption[]{The column density of C~{\sc i}  measured in individual 
components is plotted as a function of the metal depletion factor 
computed using [Fe/Zn] or [Fe/S] or [Fe/Si]. 
It is apparent from this figure that C~{\sc i} absorption line
is usually detected in components 
with high depletion factors.
}
\label{c1vsdust}
\end{figure}
\par
\subsubsection{The effect of dust}
In Fig.~\ref{c1vsdust} we plot log~$N$(C~{\sc i}) against the depletion
factor defined as log~($N$(Fe~{\sc ii})/$N$(X~{\sc ii}))$-$[Fe/X]$_\odot$
with either X~=~Zn, S or Si. 
C~{\sc i} absorption line is not detected in systems with low depletion factors
(i.e., [Fe/Zn] less than $-$0.5 dex) whereas components with higher
depletion factors readily show detectable C~{\sc i} absorption lines.
This trend is not surprising as there is a 4$\sigma$ correlation between the 
depletion factor and the metallicity of the gas in our sample (see Figure~12 
of Ledoux et al. 2003). Depletion factors lower than 0.5 dex are usually 
seen in systems with [Zn/H]~$\le -1$. High depletion factor in high metallicity 
gas implies high dust content and hence high dust optical depth to the
UV radiation. 
The absence of C~{\sc i} in components with low dust depletion
is a combination of low metallicity and low dust optical depth
to the UV radiation. It is worth remembering that similar relation exists between 
the detectability of \h2 and depletion (Fig.~14 of Ledoux et al. 2003).
\par
\subsection{C~{\sc i} fine-structure excitation}
\begin{figure}
\flushleft{\vbox{
\psfig{figure=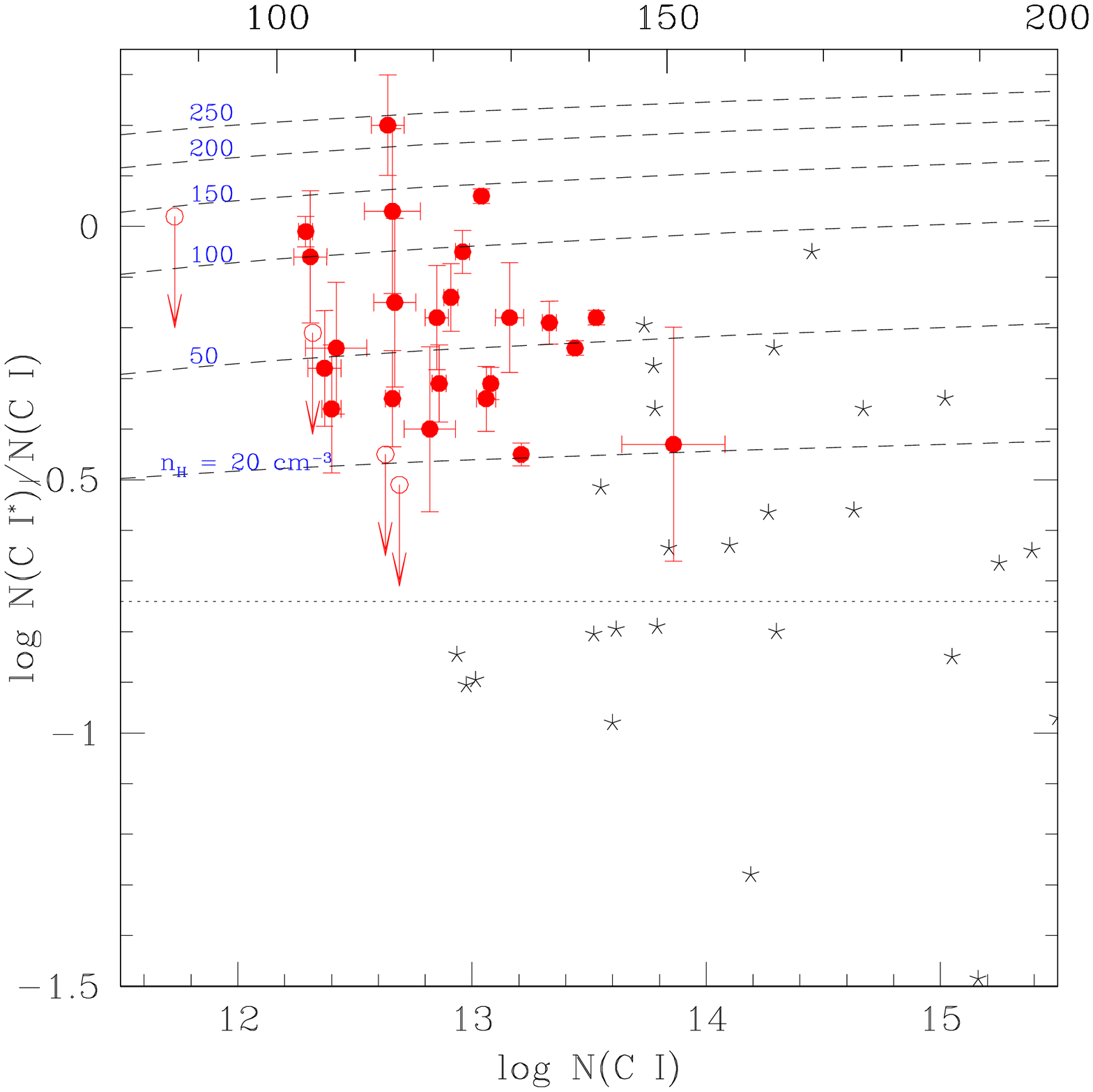,width=8.cm,height=8.cm,angle=0.}}}
\caption[]{The ratio $N$(C~{\sc i$^*$})/$N$(C~{\sc i}) is plotted 
on a logarithmic scale as a function of $N$(C~{\sc i}). 
The points with error bars are our 
measurements in individual DLA components. The stars are data measurements
in the ISM of the Galaxy drawn  from Jenkins \& Tripp (2001) and
Jenkins, Jura \& Loewenstein (1983). The horizontal
dotted line gives the expected value of the ratio if it is assumed 
that C~{\sc i} is excited by the CMBR only with $T_{\rm CMBR}$ = 8.1~K 
as expected at $z$ = 2. The short  dashed lines give the expected
ratio for different $n_{\rm H}$ as a function of temperature (top portion of
the x-axis).
}
\label{figc1}
\end{figure}
\begin{table*}
\caption{Physical conditions 
in systems in which C~{\sc i} is detected}
\begin{tabular}{llccccc}
\hline
QSO & \zabs & log $N$(S~{\sc ii}) & log $N$(C~{\sc ii}$^*$)&
\multicolumn{1}{c}{$n_{\rm e}$~(cm$^{-3}$)$^a$}&
\multicolumn{1}{c}{$n_{\rm H}$~(cm$^{-3}$)}\\
\hline
\hline
Q $0013-004$ & 1.96679$^+$ &$14.27\pm0.05$&....&$2.7~10^{-2}$&....\\
             & 1.96822$^+$ &$14.16\pm0.02$&....&$4.9~10^{-2}$&....\\
             & 1.97280$^+$ &$14.15\pm0.06$&....&$3.8~10^{-2}$&....\\
             & 1.97296$^+$ &$14.75\pm0.03$&....&$2.3~10^{-2}$&....\\
             & 1.97316$^+$ &$13.91\pm0.05$&....&$2.4~10^{-2}$&....\\
             & 1.97365$^+$ &$13.86\pm0.08$&....&$1.5~10^{-2}$&....\\
             & 1.97382$^+$ &$14.12\pm0.05$&....&$8.0~10^{-3}$&....\\
             & 1.97399$^+$ &$13.55\pm0.16$&....&$6.9~10^{-3}$&....\\
             & 1.96737     &$14.24\pm0.05$&....&$2.4~10^{-2}$&....\\
             & 1.96763     &$14.12\pm0.07$&....&$2.7~10^{-2}$&....\\
             & 1.97109     &$14.77\pm0.01$&....&$1.0~10^{-3}$&....\\
             & 1.97144     &$14.08\pm0.03$&....&$4.3~10^{-2}$&....\\ 
Q $0347-383$ & 3.02463 &$14.50\pm0.03$&$<13.00$    &$<5.0~10^{-4}$ &$<6.3$\\
             & 3.02485$^+$ &$14.36\pm0.04$&$13.55\pm0.23$&$2.3~10^{-3}$&$4.4-41.9$\\
             & 3.02501 &$13.51\pm0.12$&$<12.72$    &$<4.9~10^{-3}$ &$<32$\\
Q $0405-443$ & 2.59440     &$13.91\pm0.04$&$<12.41$    &$<3.7~10^{-3}$&$<6.3 $\\
             & 2.59464     &$14.72\pm0.02$&$13.09\pm0.10$&$<5.7~10^{-4}$&$3.0-7.5$\\
             & 2.59474$^+$ &$14.07\pm0.09$&$13.20\pm0.20$&$<2.5~10^{-3}$&$7.7-57.3$\\
             & 2.59485$^+$ &$14.73\pm0.02$&$13.25\pm0.08$&$<5.6~10^{-4}$&$4.5-10.3$\\
Q $0551-366$ & 1.96152     &$14.60\pm0.10$&$<12.00    $&$3.7~10^{-3}$&$<0.5$\\
             & 1.96168$^+$ &$14.01\pm0.27$&$13.93\pm0.09$&$3.2~10^{-2}$&$123.0-331.8$\\
             & 1.96180     &$14.23\pm0.19$&$<12.00    $&$4.6~10^{-3}$&$<1.2$\\
             & 1.96214$^+$ &$ Blended    $&$12.58\pm0.09$&$....$&$....$\\
             & 1.96221$^+$ &$ Blended    $&$13.36\pm0.07$&$....$&$....$\\
             & 1.96268     &$14.63\pm0.11$&$12.98\pm0.08$&$3.0~10^{-3}$&$2.4-9.6$\\
Q $1037-270$ & 2.13900     &$14.68\pm0.02$&$<13.50$&$2.0~10^{-3}$&$<13.2$\\
             & 2.13940     &$14.23\pm0.11$&$13.31\pm0.01$&$5.3~10^{-3}$&$<14.2$\\
Q $1232+082$ & 2.33771$^+$ &$15.24\pm0.11^b$&$<14.00    $&$<7.0~10^{-2}$&$<48$\\
Q $1444+014$ & 2.08679$^+$ &$14.32\pm0.05$&$13.12\pm0.08$&$1.1~10^{-2}$&$0.9-9.9$\\
             & 2.08692$^+$ &$14.17\pm0.06$&$12.78\pm0.20$&$1.9~10^{-2}$&$<13.0$\\
\hline
\multicolumn{6}{l}{A label ``$+$'' indicates a component in which \h2 is detected}\\
\multicolumn{6}{l}{$^a$ Derived from the Carbon ionization equilibrium}\\
\multicolumn{6}{l}{$^b$ column density of Si II is given.}\\
\end{tabular}
\label{tabphy1}
\end{table*}
In most of the DLAs with C~{\sc i} detections we also detect absorption
lines from the excited fine-structure levels.
It is therefore possible to use the relative populations of the C~{\sc i} ground
state levels to discuss the particle density, the ambient UV radiation field 
and the temperature of the cosmic-microwave background radiation 
(see Bahcall et al., 1973; Meyer et al., 1986; Songaila et al. 1994;
Ge, Bechtold \& Black, 1997; Roth \& Bauer, 1999; Srianand et al. 2000;
Silva \& Viegas, 2002; Quast et al. 2002). 
In the Galactic ISM,  fine-structure excitation of C~{\sc i}
has been used to study the distribution of thermal pressure
(see Jenkins \& Tripp 2001). 
\par
In Fig.~\ref{figc1} we plot the ratio $N$(C~{\sc i}$^*$)/$N$(C~{\sc i}) 
as a function of $N$(C~{\sc i}). 
It is clear that the C~{\sc i} column densities in DLAs (filled circles 
with error-bars) are at least an order of magnitude 
less than that measured in the ISM (stars). This is probably 
a consequence of lower metallicities and/or low H~{\sc i} content in 
DLA components.
The important point is that the $N$(C~{\sc i}$^*$)/$N$(C~{\sc i}) ratio measured in DLAs is 
remarkably larger than in the Galaxy. 
However, while comparing the ISM and DLAs, 
it is important to remember that most of the sightlines used by 
Jenkins \& Tripp (2001) have \h2 fraction orders of magnitude 
higher than what we measure in DLA components. As \h2 collisions are 
less efficient in populating the excited fine-structure state 
of C~{\sc i} we expect that for a given total
hydrogen density (and a given kinetic temperature) 
$N$(C~{\sc i$^*$})/$N$(C~{\sc i}) be higher in DLAs.
\par
The horizontal dotted line in Fig.~\ref{figc1} indicates 
the expected value of the ratio if it is assumed that the excitation 
is due to the CMBR only with a temperature $T_{\rm CMBR}$ = 8.1 K 
as expected at $z$~=~2, the typical redshift of our sample. 
It is clear that the CMBR field expected from the Big-Bang is not 
sufficient to explain the observed ratios and an extra 
contribution is required from collisional processes and/or the UV flux. 
\par
Collisions with H, He, e, p and \h2 can populate the excited
state of C~{\sc i}. 
As \hi is the dominant form of hydrogen in the gas, the contribution 
to the fine-structure excitation by \h2 collisions can be neglected. 
The electron and proton densities are expected to be very 
small, at least smaller by two orders of magnitude than the 
hydrogen density, and their contribution is also negligible 
(Keenan et al. 1986). The He~{\sc i} collisional rates are much 
less than that of H~{\sc i} and the He~{\sc i}/H~{\sc i} ratio is 
small which makes collisions with He~{\sc i} unimportant
(see Fig. 1 of Silva \& Viegas 2002). Thus in our analysis of 
the C~{\sc i} excitation we consider only collisions by neutral 
hydrogen. The rates are taken from  Launay \& Roueff (1977). 
The spontaneous decay rates are 
${\rm A_{10} = 7.93\times 10^{-8}~s^{-1}}$  and 
${\rm A_{21} = 2.68\times 10^{-7}~s^{-1}}$ 
(Bahcall \& Wolf 1968). The corresponding 
CMBR excitation rate is derived from these values.
The UV pumping rate in the cloud depends on the nature and strength 
of the UV radiation field. We assume that the UV intensity is the 
same as in the ISM of our Galaxy as suggested by the high$-J$
excitation of \h2.
\par
The dashed lines in Fig.~\ref{figc1} give the expected ratio for, 
$T_{\rm CMBR}$ = 8 K, a UV radiation field like in the Galaxy 
(with an excitation rate of 
$\Gamma_{01} = 7.55\times10^{-10}~{\rm s}^{-1}$), hydrogen 
density in the range $n_{\rm H}$~=~20$-$250~cm$^{-3}$, 
versus the kinetic temperature, $T_{\rm kin}$, in the range 
$80-200$ K (consistent with $T_{01}$ measured in DLAs).
It can be seen that a typical density range consistent with most 
of the observed point is, 20$\le n_{\rm H}{\rm (cm^{-3})}\le $150. 
The density ($n_{\rm H}$) and pressure ($p/K$) derived for
individual components are summarized in column 9 and 10
respectively in Table.~\ref{tabphy}.
Here, we assume $T_{\rm CMBR}$~=~2.7$\times$(1+\zabs) and 
$T$~=~$T_{01}$ in the case of \h2 detection and $T$~=~100 K otherwise. 
The derived pressure range in DLA components are higher than that 
typically measured in the galactic ISM (see Jenkins \& Tripp 2001)
and consistent with what is expected in a the cold neutral
medium (CNM) with lower metallicity (Z$\sim$0.1 Z$_\odot$) 
and dust depletion (see Wolfire et al. (1995, 2001), Wolfe et al., (2003a, b)
Srianand et al. 2005).
Interestingly the derived density range in most of the components is
close to the critical density for thermalising the \h2 $N$(J=2)/$N$(J=0) 
ratio (see Fig.~\ref{h2col}).
\section{Carbon ionization state}
\begin{figure}
\flushleft{\vbox{
\psfig{figure=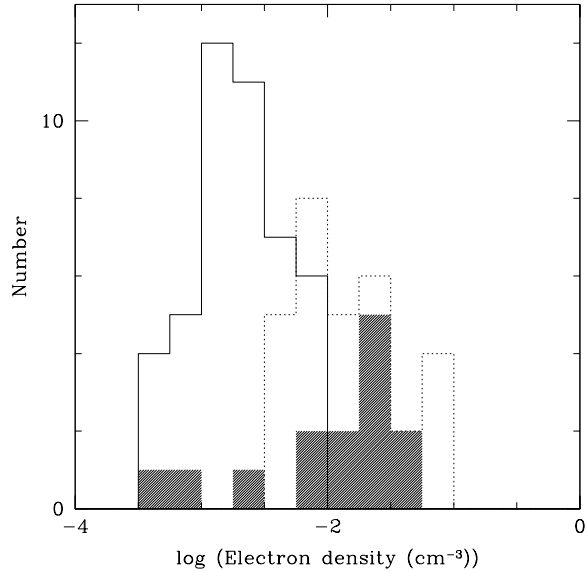,width=8.cm,height=8.cm,angle=0.}}}
\caption[]{Histogram of the electron density derived using Eq.~\ref{ionz}
for $T$ = 100 K. The shaded histogram is based on systems that 
show detectable C~{\sc i} absorption lines. The histograms with 
continuous and dotted lines are upper 
limits on the electron density for, respectively, systems with 
log~$N$(C~{\sc ii})
greater and less than 15. We use the C~{\sc i} photoionization rate 
given for the Galactic UV field by P\'equignot \& Aldrovandi (1986).
}
\label{eden}
\end{figure}
In the ISM, when hydrogen is neutral, most of the carbon is in the form 
of C~{\sc ii} so that we can use the C~{\sc i}/C~{\sc ii} ratio to derive the
physical conditions in the gas. 
As already mentioned, the C~{\sc ii} absorption features are always badly
saturated and we rely on the realistic assumption that C~{\sc ii} in the 
neutral phase can be traced by S~{\sc ii} and/or Si~{\sc ii} so that we can 
use the weak lines of these species to derive the C~{\sc ii} column densities
in the components of interest. 
\par
Assuming photoionization equilibrium between C~{\sc i} and C~{\sc ii} 
and using the atomic data from Shull \& Van Steenberg (1982)
and P\'equignot et al. (1991), we can write,
\begin{equation}
{n_{\rm e}\over \Gamma} =  4.35\times10^{11}{N({\rm C~{I}})\over N({\rm C~{II}})}
\bigg{(} {T\over 10^{4}}\bigg{)}^{0.64}
\label{ionz}
\end{equation}
\noindent
Here, $\Gamma$ is the photoionization rate for C~{\sc i}.  In the local ISM,
$\Gamma_{\rm gal}\simeq2-3.3\times10^{-10}$~s$^{-1}$ 
(P\'equignot \& Aldrovandi 1986). Here, we neglect the ion-molecular
interaction and charge exchange reactions that may produce C~{\sc i}.
Thus $n_{\rm e}/\Gamma$ can be constrained once the temperature of the gas is known.
Application of Eq.~\ref{ionz} to the cold Galactic ISM 
(with $T$ = 100 K) has resulted in $n_{\rm e}\simeq0.14\pm0.07$ cm$^{-3}$ 
(Welty et al. 2002 and references there in). In the 
stable CNM considered by Wolfire et al. (1995) the electron density is expected
to be in the range 0.01$\le n_{\rm e} ({\rm cm^{-3}})\le$0.02 for $Z$ = 0.1~$Z_\odot$
and dust abundance one tenth of the Galactic ISM.
\subsection{Systems with C~{\sc i} detections}
First we concentrate on the systems with C~{\sc i} detections.
We estimate the electron density assuming a UV field similar to the 
Galactic mean field (i.e., $\Gamma_{\rm gal} = 2.5\times10^{-10}$~s$^{-1}$), 
$T = T_{01}$ for 
the \h2 components and $T$ =100 K otherwise.
Individual values of $n_{\rm e}$ derived for these systems are given in
Table~\ref{tabphy1}.
The electron density is in the range 
0.7$\times10^{-2}\le n_{\rm e}~({\rm cm^{-3})} \le 4.9\times10^{-2}$. 
Together with \nh given in Table~\ref{tabphy}, this suggests that 
$n_{\rm e}/n_{\rm H}\le 10^{-3}$ for most of the systems. 
Therefore, 
the ionization state of the gas with \h2 and C~{\sc i} is similar to that in 
the CNM in a moderate radiation field.
\subsection{Systems without \h2 and C~{\sc i}}
In the case of systems in which neither \h2 nor C~{\sc i} are detected,
we assume $T$~=~100~K and $\Gamma_{\rm gal}=2.5\times10^{-10}$~s$^{-1}$ 
and obtain upper limits on the electron density. 
The results are plotted 
in Fig.~\ref{eden}. We notice that the inferred electron densities are
much smaller than in systems in which C~{\sc i} (and \h2) are detected.
The difference is even larger if we use only the systems with 
log~$N$(C~{\sc ii})$\le$15.
For this gas the inferred electron densities are less than
10$^{-2}$ cm$^{-3}$ (with a median of 10$^{-3}$ cm$^{-3}$).
This may indicate that the absorption
originates from warm neutral medium (say $T$ = 8000 K). 
If the average radiation field in DLAs is similar to that in the Galactic
ISM then the absence of C~{\sc i} in most of the DLAs could just be a 
consequence of lower densities (and/or higher $T$) in these systems. 
One can derive an independent estimate of the particle density using 
the excitation of C~{\sc ii} fine structure levels. This is what we do in 
the following Section.
\par
\section{Excited fine-structure line of C~{\sc ii}}
\begin{figure*}
\flushleft{\vbox{
\psfig{figure=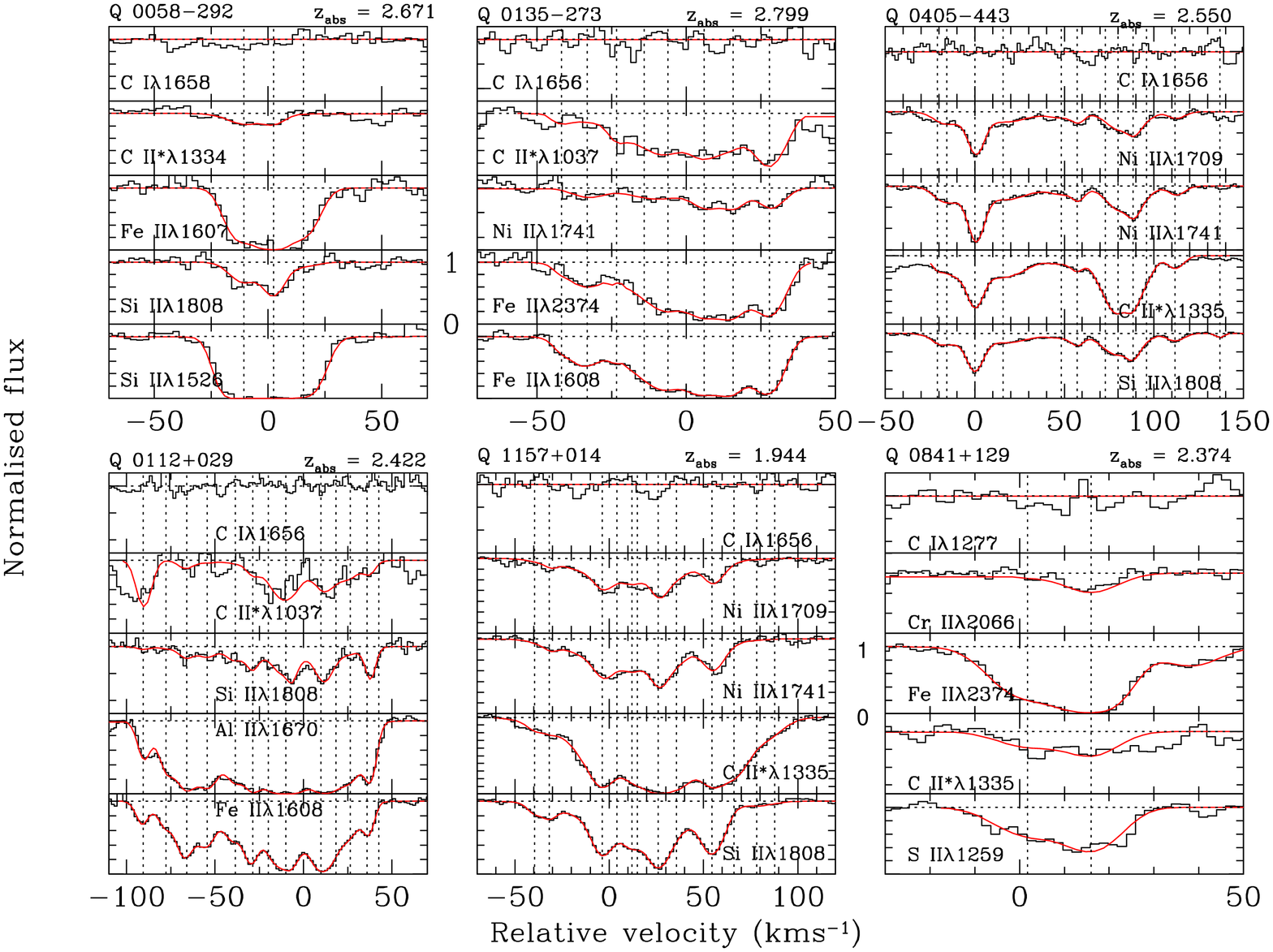,width=16.cm,height=11.cm,angle=270.}}}
\caption[]{Voigt profile fits to systems with C~{\sc ii$^*$}
absorption line and without \h2 or C~{\sc i} absorption lines. In each panel,
the histogram gives the observed spectrum and the continuous curve
is the best fitted multicomponent Voigt profile. The vertical dotted lines 
mark the locations of individual components.
}
\label{vpc2star}
\end{figure*}
\begin{table*}
\caption {Results of Voigt profile fits to DLAs that do not show detectable \h2 absorption lines}
\vskip -0.1in
{\tiny
\begin{tabular}{lcccccccclc}
\hline
Quasar&log $N$(H~{\sc i}) & $Z/Z_\odot$&\zabs & $\log N($C\,{\sc i}$)$ & $\log N($Si\,{\sc ii}$)$ & $\log N($S\,{\sc ii}$)$ & [Fe/X] & 
$\log N($C\,{\sc ii}\,$^\star)$& $b$(km s$^{-1}$)& X\\
\hline
\hline
Q\,0000$-$263 & 21.40(0.08)&  $-2.06$(0.09)&  3.39013&....&15.06(0.02)& 
14.70(0.03)  &$+0.03$(0.06)&....&10.0(0.3) & Zn\\
Q\,0010$-$002&20.95(0.10)&$-1.43$(0.11)&2.02471&$<12.15$&14.90(0.05)&14.50(0.05)&$+0.18$(0.08)&$<13.40$& 5.5(0.4)&Zn\\
             &           &             &2.02484&$<12.15$&15.09(0.03)&14.77(0.03)&$+0.15$(0.07)&$<13.40$&8.4(0.5)&Zn\\
Q\,0058$-$292& 21.10(0.10)&$-1.53$(0.10)& 2.67123&$<12.35$&14.87(0.07)&14.61(0.03)&$-0.69$(0.05)&12.82(0.04)&8.7(0.7)& Zn\\
             &            &             & 2.67142& $<12.35$&14.98(0.05)&14.63(0.03)&$-0.13$(0.04)&12.79(0.04)& 8.2(0.2)&  Zn\\
Q\,0102$-$190 &21.00(0.08)&$-1.90$(0.09)&2.36958 &$<12.30$&....&        
13.93(0.04)&$-0.16$(0.06) &$<12.92^+$  &2.0(0.2)&  S\\
              &           &             &2.36966 &$<12.30$&....&14.06(0.03)&
$-0.17$(0.04)&$<13.02^+$&3.5(0.2)&  S\\
Q\,0112$-$306 &20.50(0.08) &$-2.43$(0.09)&2.41844&$<12.32$&13.31(0.02)& 
$<14.14$&$-0.15$(0.04)&$<13.20$&3.5(0.2)&Si\\
              &            &             &2.41861&$<12.32$&13.33(0.02)& 
$<14.14$&$-0.33$(0.09)&$<13.20$&6.9(0.4)&Si\\
              &            &             &2.41869&$<12.32$&$<12.44$&    
$<14.14$&$>+0.33$&$<13.20$&5.6(1.1)& Si\\
              &20.30(0.10) &$-0.50$(0.15)&2.70111&$<12.20$&$<14.20$&
.... &$>-0.61$&$<13.20$&7.2(0.8)&  Si\\
              &            &             &2.70163&$<12.20$&14.75(0.05)& 
....&$-0.48$(0.06)&$<13.00$&27.4(2.2)&  Si\\
              &            &             &2.70217&$<12.20$&14.76(0.10)&  
....&$-0.66$(0.15)&$<13.00$&14.9(3.1)&  Si\\
              &            &             &2.70233&$<12.20$&14.29(0.18)&  
....&$-0.21$(0.22)&$12.60\pm0.08$&3.6(1.2)&  Si\\
              &            &             &2.70257&$<12.20$&14.58(0.05)& 
....& $-0.47$(0.06)&$12.97\pm0.06$&11.0(1.2)&  Si\\
              &            &             &2.70276&$<12.20$&$<14.20$&   
....&$>-0.43$ &$<12.60$&3.8(0.7)&  Si\\

             &            &             &2.70316&$<12.20$&$13.82(0.01)$&     
....&$-0.40(0.06)$ &12.75$\pm$0.06&7.2(1.0)&  Si\\
              &            &             &2.70332&$<12.20$&14.18(0.03)&  
....&$-0.66$(0.10)&13.20$\pm$0.04&4.6(0.5)&  Si\\
              &            &            &2.70334&$<12.20$&14.21(0.01) &
    &$-1.00(0.10)$&13.71$\pm$0.02&11.8(1.1)&Si\\
Q\,0112$+$029&20.90(0.10)&$-$1.32(0.15)&2.42234&$<11.78$&13.92(0.13)&....&$-$0.18(0.13)&12.56(0.24)&4.7(0.3)&S\\
            &            &&2.42248&$<11.78$&14.04(0.11)&....&$-$0.35(0.12)&12.14(0.72)&6.8(0.8)&S\\ 
            &            &&2.42266&$<11.78$&14.02(0.10)&....&$-$0.58(0.11)&12.21(0.59)&6.1(0.9)&S\\
            &            &&2.42277&$<11.78$&14.16(0.08)&....&$-$0.37(0.09)&12.70(0.17)&3.7(0.3)&S\\
            &            &&2.42298&$<11.78$&14.77(0.04)&....&$-$0.40(0.05)&13.66(0.06)&10.5(0.5)&S\\
            &            &&2.42323&$<11.78$&14.65(0.04)&....&$-$0.38(0.05)&13.26(0.07)&6.3(0.3)&S\\
            &            &&2.42338&$<11.78$&14.04(0.14)&....&$-$0.33(0.15)&13.20(0.08)&9.5(1.3)&S\\
            &            &&2.42353&$<12.22$&13.97(0.06)&....&$-$1.13(0.09)&12.61(0.20)&2.5(0.6)&S\\
Q\,0135$-$273&21.00(0.10) &$-$1.40(0.10)&2.79948&$<$12.00&13.91(0.05)&....&$-$0.27(0.07)&12.78(0.25)&7.4(1.1)&S\\
            &            &&2.79960&$<$12.00&14.14(0.10)&....&$-$0.33(0.12)&13.50(0.09)&0.1(0.1)&S\\ 
            &            &&2.79972&$<$12.00&14.14(0.31)&....&$-$0.24(0.32)&13.39(0.12)&8.4(1.4)&S\\ 
            &            &&2.79982&$<$12.00&14.08(0.30)&....&$-$0.24(0.32)&13.38(0.12)&5.5(1.0)&S\\
            &            &&2.79998&$<$12.00&13.85(0.10)$^*$&....&$ $0.39(0.11)&13.69(0.06)&6.2(0.8)&S\\
            &            &&2.80010&$<$12.00&14.05(0.38)$^*$&....&$ $0.08(0.38)&13.26(0.10)&3.9(0.4)&S\\
            &            &&2.80025&$<$12.00&14.71(0.08)&....&$-$0.53(0.08)&13.85(0.07)&5.1(0.2)&S\\
Q\,0405$-$443&21.15(0.15)&$-$1.41(0.15)&2.54967&$<11.70$&12.99(0.27)&....&$-$0.84(0.82)&11.83(0.29)&0.3(0.1)&S\\
            &            &&2.54973&$<11.70$&13.23(0.03)&....&$-$0.19(0.04)&12.75(0.05)&9.8(0.9)&S\\
            &            &&2.54992&$<11.70$&14.56(0.02)&....&$-$0.23(0.02)&13.12(0.03)&5.3(0.2)&S\\
            &            &&2.55011&$<11.70$&14.41(0.05)&....&$-$0.19(0.05)&13.21(0.04)&18.5(2.1)&S\\
            &            &&2.55049&$<11.70$&13.76(0.11)&....&$-$0.09(0.12)&12.45(0.12)&9.4(2.2)&S\\
            &            &&2.55060&$<11.70$&13.87(0.05)&....&$-$0.31(0.06)&12.38(0.07)&3.8(0.6)&S\\
            &            &&2.55078&$<11.70$&14.10(0.05)&....&$-$0.48(0.06)&12.96(0.09)&7.8(1.3)&S\\
            &            &&2.55087&$<11.70$&14.18(0.12)&....&$-$0.26(0.12)&13.16(0.12)&6.2(1.4)&S\\
            &            &&2.55097&$<11.70$&14.30(0.06)&....&$-$0.31(0.06)&13.08(0.09)&4.5(0.4)&S\\
            &            &&2.55105&$<11.70$&13.69(0.05)&....&$-$0.19(0.07)&12.75(0.03)&5.1(0.5)&S\\
            &            &&2.55124&$<11.70$&13.86(0.03)&....&$-$0.29(0.04)&12.55(0.03)&5.9(0.5)&S\\

            &20.45(0.10) &$-2.04$(0.11) & 2.62101&  $<11.64$  &   12.92(0.02)& 
$<13.37$    & $-0.23$(0.04) &$<11.75$   &7.5(0.2)&  Si\\
            &            &              &2.62140 & $<11.64$   &  13.20(0.03)& 
$<13.37$    & $-0.23$(0.04)&  $<11.75$& 5.4(0.2) & Si\\
            &             &              &2.62158 & $<11.64$   &  13.47(0.02)& 
$<13.37$    & $-0.28$(0.03)&  $<11.75$& 16.0(0.6)&  Si\\
            &             &              &2.62178 &$<11.64$    & 12.92(0.02)& 
$<13.67$    & $-0.22$(0.04)&  $<11.75$& 4.4(0.2) & Si\\
            &             &              &2.62219 & $<11.64$   & 12.02(0.05)& 
$<13.67$    & $<+0.18$    &   $<11.75$& 5.7(1.3) & Si\\
            &             &              &2.62235 & $<11.64$   & 11.87(0.06)& 
$<13.67$    & $<+0.33$    &   $<11.75$   &2.0(1.3)&  Si\\
            &             &              & 2.62287&  $<11.64$  & 12.98(0.02)& 
$<13.24$    & $-0.55$(0.06)&  $<11.75$&10.9(0.5)&  Si\\
            &             &              &2.62311&  $<11.64$&     13.03(0.08)& 
$<13.24$    & $-0.55$(0.12)&  $<11.75$&6.9(0.9)&  Si\\
            &              &             & 2.62320 & $<11.64$   &  12.96(0.08)& 
$<13.24$    & $-0.55$(0.13)&  $<11.75$&3.6(0.3)&  Si\\
Q\,0841$+$129&21.05(0.10)&$-$1.59(0.10)&2.37439&$<12.21$&....&14.26(0.18)&$-$1.14(0.20)&12.91(0.14)&9.0(2.3)&S\\ 

            &            &             &2.37455&$<12.21$&....&14.43(0.12)&$-$0.60(0.23)&13.01(0.11)&7.2(1.3)&S\\
            &20.80(0.10)&$-$1.60(0.10)&2.47604&$<12.21$&....&13.47(0.27)&$-$0.51(0.34)&$<12.70$&5.1(2.4)&S\\
            &            &&2.47621&$<12.21$&....&14.42(0.05)&$-$0.26(0.07)&$<12.70$&9.1(1.1)&S\\
            &            &&2.47642&$<12.21$&....&13.35(0.27)&$-$0.40(0.30)&$<12.70$&2.2(2.5)&S\\

%
Q\,1037$-$270&19.70(0.05) &$-0.31$(0.06)&2.13910&12.51(0.02)&15.10(0.02)& 
14.65(0.01)&$-0.20$(0.03)&$< 13.50$&15.6(0.2)&  Zn\\
             &            &             &2.13951&12.46(0.02)&14.70(0.02)& 
14.23(0.01)&$-0.05$(0.04)&$< 13.08$&8.8(0.2)&  Zn\\

Q\,1101$-$264&19.50(0.04)&$-1.07$(0.06)&1.83817 &$<11.68$ &12.29(0.09)& 
$<13.05$&$-0.85$(0.18)&$<11.83$ &4.5(0.4)&Si\\
             &           &             &1.83831 &$<11.68$ &12.68(0.05)& 
$<13.05$&$-0.75$(0.09)&$<11.83$ &10.3(0.3)&Si\\
             &           &             &1.83854 &$<11.68$ &13.33(0.02)& 
13.19(0.26)&$-0.49$(0.02)&$<11.83$&5.8(0.1)&Si\\
             &           &             &1.83871 &$<11.68$ &13.21(0.02)& 
13.12(0.36)& $-0.49$(0.03)&$<11.83$&9.4(0.2)&Si\\
             &           &             &1.83890 &$<11.68$ &13.61(0.02)& 
13.32(0.19)&$-0.45$(0.02)& $<11.83$&5.4(0.1)&Si\\
             &           &             &1.83915 &$<11.68$ &12.77(0.05)& 
$<13.05$ &$-0.23$(0.05)&b&5.6(0.2)& Si\\
             &           &             &1.83925 &$<11.68$ &12.50(0.28)& 
$<13.05$ &$-0.15$(0.28)&b&2.0(0.2)& Si\\
             &           &             &1.83933 &$<11.68$ &12.44(0.26)& 
$<13.05$ &$-0.42$(0.26)&b &5.0(0.5)&  Si\\
Q\,1117$-$134 &20.95(0.10) &$-1.42$(0.14)&3.35027 &$<12.76$&14.78(0.03)& b&$-0.27$(0.06)&$< 13.66$&6.9(0.4)&Zn\\
              &            &             &3.35046 &$<12.45$&14.68(0.05)&b&$-0.21$(0.06)& $< 13.31$&6.5(0.6)&Zn\\
              &            &             &3.35067 &$<12.45$&14.26(0.05)&b&$>-0.47$&$< 13.29$& 6.2(0.3)&  Zn\\
Q\,1157$+$014& 21.80(0.10)&$-$1.40(0.10)&1.94311&$<12.15$&14.13(0.08)&....&$-$1.40(0.30)&12.36(0.14)&3.9(1.3)&Si\\
            &            &&1.94318&$<12.15$&14.22(0.08)&....&$-$0.96(0.13)&12.21(0.23)&3.6(1.1)&Si\\
            &            &&1.94354&$<12.15$&15.32(0.04)&....&$-$0.67(0.05)&13.91(0.04)&28.9(1.4)&Si\\
            &            &&1.94346&$<12.15$&15.09(0.03)&....&$-$1.16(0.06)&13.74(0.04)&6.0(0.4)&Si\\
            &            &&1.94361&$<12.15$&14.75(0.23)&....&$-$1.10(0.31)&13.00(1.37)&6.1(1.5)&Si\\
            &            &&1.94375&$<12.15$&15.39(0.03)&....&$-$1.21(0.04)&14.08(0.18)&5.6(0.0)&Si\\
	    &            &&1.94403&$<12.15$&15.26(0.01)&....&$-$1.07(0.02)&13.87(0.02)&7.2(0.0)&Si\\
            &            &&1.94385&$<12.15$&14.89(0.03)&....&$-$0.92(0.05)&13.96(0.03)&8.1(0.5)&Si\\ 
	    &            &&1.94426&$<12.15$&14.07(0.69)&....&....&13.14(0.06)&10.3(0.0)&Si\\
Q\,1223$+$178 &21.40(0.10) &$-1.63$(0.11)& 2.46530 & $<12.30$& 14.90(0.02)&14.56(0.03)&  $-0.12$(0.04)& 13.63(0.02)&10.3(0.3)&Zn\\
              &            &             & 2.46559 & $<12.30$& 14.78(0.03)&14.54(0.04)&  $+0.02$(0.06)&$< 13.42$ &14.0(0.8)&Zn\\
              &            &             & 2.46607 & $<12.30$& 15.22(0.02)&14.82(0.02)&  $-0.05$(0.03) & $< 13.04$&14.7(0.3)&Zn\\
              &            &             & 2.46628 & $<12.30$& 14.13(0.09)& $<13.85$ &    ....         &  $<12.55$& 4.3(0.7)&  Zn\\


Q\,1337$+$113&20.12(0.05) & $-1.81$(0.07)&2.50766  &$<12.45$ &12.93(0.06)&  b& $-0.58$(0.09)&$<12.55$&5.0(1.3)&  Si\\
             &            &              &2.50792  &$<12.45$ &13.82(0.04)&  b& $-0.40$(0.04)& $<12.55$&6.5(0.2)&  Si\\

            &21.00(0.08) &$-1.86$(0.10) & 2.79557  &$<12.40$ &13.90(0.16)&  ....& $-0.22$(0.17)&b&12.8(1.6)&Si\\
            &             &              & 2.79584  &$<12.40$ &14.63(0.04)&  ....& $-0.34$(0.04)&  b & 6.6(0.3)&Si\\


Q\,1451$+$123 &20.40(0.10)&$-2.27$(0.14) &2.46897&$<12.75$ &13.23(0.09)&$<13.55$&$-0.21$(0.09)&  b&6.7(0.8)&Si\\
              &           &              &2.46921&$<12.75$ &13.50(0.11)&$<13.55$&$-0.26$(0.11)&  b &5.7(0.5)&  Si\\

              & 20.20(0.20)&$-2.10$(0.21)&3.17081&$<12.65$&13.31(0.06)&  ....& $-0.29$(0.14) & $<13.00$ &9.9(1.8) & Si\\
              &            &             &3.17112&$<12.65$&13.31(0.07)&  ....&$-0.27$(0.13)  &$<13.00$ & 7.3(1.5) & Si\\
              &            &             &3.17142&$<12.65$&12.72(0.18)&  ....&$<+0.19$       &$<13.00$ &9.6(4.9)  &Si\\
Q\,2059$-$360 &20.29(0.07)&$-1.94$(0.08)& 2.50734&$<12.25$&13.85(0.03)&13.49(0.23)&$-0.36$(0.04)&$<12.53$&4.5(0.3) &Si\\
              &           &             & 2.50753&$<12.25$&12.99(0.10)&$<13.40$   &$-0.29$(0.13)&$<12.53$&5.8(6.1) &Si\\

              &20.98(0.08) &$-1.78$(0.11)& 3.08261&$<12.10$&14.36(0.05)&13.97(0.05)&$-0.29$(0.05)& $<12.59$&5.7(0.3)&S\\
              &            &             & 3.08291&$<12.10$&14.61(0.05)&14.16(0.05)&$-0.25$(0.05)& $<12.59$&8.7(0.6)&S\\
              &            &             &3.08314 &$<12.10$&$\le 14.27$& $<13.56$ &    $>-0.19$ &$<12.59$&7.2(0.9)  &S\\

Q\,2138$-$444 &20.98(0.05)&$-1.74$(0.05)&2.85234&$<12.02$  &14.86(0.02)&14.50(0.02)&$-0.11$(0.03)&$<13.09$&8.5(0.2)&  Zn\\
%
Q\,2332$-$094& 20.50(0.07)&$-1.33$(0.11)&3.05632&$<12.02$&sat/bld&13.76(0.05)&$-0.50$(0.05)&$<12.52$&9.8(0.3)&  S\\
             &            &             &3.05657&$<12.02$&sat/bld&13.49(0.07)&$-0.51$(0.08)&$<12.52$&5.1(0.5)&  S\\
             &            &             &3.05676&$<12.02$&sat/bld&$<13.42$   &$>-0.44$     &$<12.52$&3.2(0.4)&  S\\
             &            &             &3.05690&$<12.02$&sat/bld&13.48(0.11)&$-0.09$(0.11)&$<12.52$&12.9(0.6)& S\\
             &            &             &3.05722&$<12.02$&sat/bld&13.85(0.04)&$-0.18$(0.04)&$<12.52$&4.9(0.2)& S\\
             &            &             &3.05738&$<12.02$&sat/bld&13.50(0.09)&$-0.55$(0.10)&$<12.52$&6.6(0.7)& S\\
\hline
\multicolumn{11}{l}{The numbers in parentheses are $1\sigma$ standard deviations. Strict upper limits are $5\sigma$ detection limits;``....'':  no transition line from this ion has been observed }\\
\multicolumn{11}{l}{``b'': blended; ``sat/bld'': most or all of the components in the observed transition
lines from this ion are either saturated or blended, or both. }\\
\end{tabular}
}
\label{tabphy2}
\end{table*}
\subsection{Method to derive physical parameters}
Under LTE, the column density ratio $N$(C~{\sc ii$^*$})/$N$(C~{\sc ii}), can be written as,
\\  
\begin{equation}
{N({\rm C~{II^*}})\over N({\rm C~{\sc II}})} = {Q_{12}(\rm e)~n_{\rm e} +Q_{12}(H)~n_{\rm H}+\Gamma_{12}({\rm CMB})\over A_{21}}
\label{c2seq}
\\
\end{equation}
\par\noindent
where, 
$Q_{12}({\rm e}) = 7.8 \times 10^{-6} exp[-91.27/T] ~T^{-0.5}$ cm$^{-3}$ s$^{-1}$ and
$Q_{12}({\rm H}) = 1.3 \times 10^{-9} exp[-91.27/T]$ cm$^{-3}$ s$^{-1}$ are the collisional
excitation rates per unit volume for electrons and hydrogen atoms (Bahcall \& Wolf 1968) 
with $T$ being the kinetic temperature of the gas. The Einstein's coefficient is 
$A_{21} = 2.291\times10^{-6}$ s$^{-1}$.
The CMB pumping rate, $\Gamma_{12}{\rm (CMB)}$, equals 6.6$\times10^{-11}$ s$^{-1}$ and
1.1$\times10^{-9}$ s$^{-1}$ for, respectively, redshifts 2 and 3 (Silva \& Viegas 2002). 
For a given temperature, the collisional excitation rate for electrons is orders of magnitude 
larger than that for hydrogen atoms. For example, when $T$ = 1000 K,
whenever $n_{\rm e}/n_{\rm H}$ is larger than 5$\times10^{-3}$, collisions with 
electrons is the dominant process. UV pumping is an additional possible excitation 
mechanism. For the mean radiation field in our Galaxy, the UV pumping rate is 
9.3$\times10^{-11}$ s$^{-1}$ (Silva \& Viegas 2002). This is similar
to or slightly lower than the CMB pumping rate for the range of redshift
we consider in this study.
\par
We compute the expected value of log~$N$(C~{\sc ii}$^*$)/$N$(C~{\sc ii}) ratio
under different situations and in particular the WNM and CNM solutions given in 
Table~3 of Wolfire et al. (1995). Note that if CMBR  pumping alone is responsible 
for the excitation, the expected ratios 
are $-4.54$ and $-3.31$ for, respectively, \zabs = 2 and 3. 
For the standard ISM (with stable pressure in the range 990$-$3600 cm$^{-3}$ K), 
we derive $-2.62\le$log~$N$(C~{\sc ii$^*$})/$N$(C~{\sc ii})$\le -2.20$ 
for the CNM and $-3.65\le$log~$N$(C~{\sc ii$^*$})/$N$(C~{\sc ii})$\le -3.17$ for 
the WNM. As the DLA gas has low metallicity and low dust content,
the expected pressure should be higher in the two DLA phases (see Liszt 2002
and Wolfe et al. 2003). If we assume $Z = 0.1 Z_\odot$ and dust to
gas ratio one tenth of the ISM value (which is typical of DLAs) then we expect 
$-2.26\le$log~$N$(C~{\sc ii$^*$})/$N$(C~{\sc ii})$\le -2.08$ 
for the CNM and $-3.39\le$log~$N$(C~{\sc ii$^*$})/$N$(C~{\sc ii})$\le -2.70$
for the WNM in DLAs. Thus if DLAs originate from H~{\sc i} gas in a two-phase 
equilibrium, we expect the CMBR pumping to be sub-dominant compared to 
collisional excitation. In fact if the gas is completely neutral 
then  from Eqs. (\ref{ionz}) and (\ref{c2seq}) we derive,
\begin{eqnarray}
n_{\rm H} &=&{\rm {A_{21}N(C~{II^*})\over\sigma_H N(C~II)} - 
{\sigma_e n_e\over \sigma_H }}\nonumber\\
    &=& {\rm {A_{21}N(C~{II^*})\over\sigma_H N(C~II)} - 
{\sigma_e N(C~I)\Gamma \over \sigma_H  \alpha_r N(C~II)}}.
\label{enheq}
\end{eqnarray}
This gives an independent estimate of \nh which, by comparison with the estimate
derived from the C~{\sc i} excitation, can lead to constraints on the radiation field. 
However, the ionization fraction (i.e $n_{\rm e}/n_{\rm H}$) and the temperature of 
the neutral absorbing gas must be accurately determined before densities can be 
derived using  the $N$(C~{\sc ~ii}$^*$)/$N$(C~{\sc ii}) ratio.
Indeed, the fact that, in DLAs,  the Al~{\sc iii} absorption profile is very similar to that
of neutral or singly ionized species has been used as evidence for the presence of 
ionized gas being mixed with the neutral gas in DLAs
(Lu et al. 1996; Prochaska \& Wolfe 1999; Howk \& Sembach 1999; 
Wolfe \& Prochaska 2000; Vladilo et al. 2001; Izotov et al. 2001).
In the ionized gas (i) $n_{\rm e}$ as well as $T$ will be higher than what is expected 
in the warm or cold neutral gas and (ii) $N$(Si~{\sc ii}) will under-predict 
$N$(C~{\sc ii}) as the ionization corrections are 
different for the two species (see Fig.~1 of Izotov et al. 2001). Neglecting
the presence of ionized gas can artificially enhance the derived \nh~values.
\subsection{Frequency of C~{\sc ii}$^*$ detection}
In our sample, all the systems that show \h2 also show detectable C~{\sc ii$^*$} 
absorption line. The details of the fits to the C~{\sc ii$^*$} absorption 
line for these 
systems are summarized in Table~\ref{tabphy2} and shown in Fig.~\ref{vpc2star}. 
The components toward Q\,$0013-004$ and Q\,$0528-250$ are badly blended and it is 
therefore not possible to fit $N$(C~{\sc ii$^*$}).
We detect C~{\sc ii$^*$} in seven out of the 21 DLAs that do not show \h2 absorption lines.
If we also include the 8 DLAs that show \h2, about 50\% (15 out of 29) of the DLAs in 
our sample show detectable C~{\sc ii$^*$} absorption line 
which is consistent with the
finding by Wolfe et al. (2003).
\begin{figure}
\flushleft{\vbox{
\psfig{figure=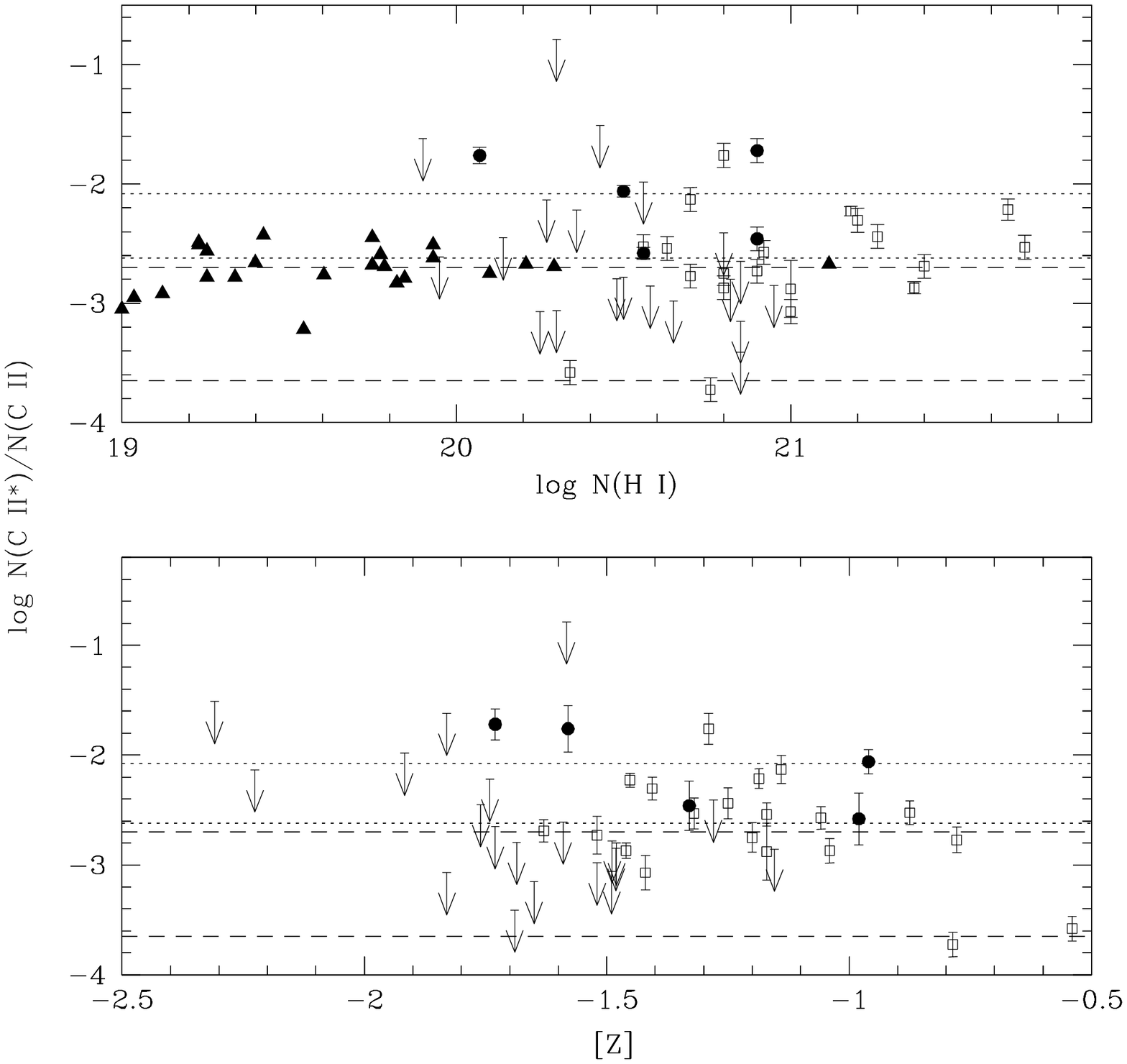,width=8.cm,height=8.cm,angle=0.}}}
\caption[]{The  average (over the whole profile) $N$(C~{\sc ii$^*$})/$N$(C~{\sc ii}) 
ratio in DLAs (combining our sample to measures by Wolfe et al., 2003a, 2003b) is plotted
against log~$N$(H~{\sc i}) (upper panel) and the metallicity Z 
(lower panel).
The dotted lines show the expected range of values observed in the CNM gas 
with metallicity and dust-content ranging from that observed in the
Galactic ISM to that of a typical DLA. The dashed lines give
the corresponding range for the WNM gas. 
 Filled circles and open squares are respectively
for systems with and without \h2 detections.
In the top panel the filled triangles are the observations of 
Lehner et al. (2004) 
along high latitude Galactic sightlines with low \h2 content. 
Most of the points 
with log $N$(H~{\sc i}) $\le20.0$  are from intermediate or high velocity
clouds in the Galactic halo.
}
\label{c2sdet}
\end{figure}
\par
In Fig.~\ref{c2sdet} we plot the  average (over the whole profile) 
$N$(C~{\sc ii$^*$})/$N$(C~{\sc ii}) ratio measured in DLAs in our sample 
together with the measurements by Wolfe et al. (2003a, 2003b)
against the total H~{\sc i} column density and the silicon metallicity. 
Here we use the total column density summed over all the components. 
$N$(C~{\sc ii}) is computed from $N$(Si~{\sc ii}) assuming
solar abundance ratio without any ionization correction.
From the upper panel in Fig.~\ref{c2sdet} it can be seen that C~{\sc ii$^*$} 
is detected in all the systems with log $N$(H~{\sc i}) $\ge$21.0. Interestingly 
no such relationship exists between C~{\sc i} (or \h2) and H~{\sc i}.
Most the systems with  log $N$(H~{\sc i}) $\ge$21.0 have 
$N$(C~{\sc ii$^*$})/$N$(C~{\sc ii}) consistent with what is expected
in CNM. On the contrary, the measured values of $N$(C~{\sc ii$^*$})/$N$(C~{\sc ii}) in
systems with lower $N$(H~{\sc i}) spread over more than an order of magnitude
covering the expected ranges for WNM and CNM. 

From the bottom panel, it can be seen that C~{\sc ii$^*$} is frequently 
detected in 
gas with high metallicity as already noticed by Wolfe et al. (2003a). 
Most of the systems that show C~{\sc ii$^*$} absorption line with 
lower $N$(H~{\sc i}) do have statistically higher metallicity.
In the whole sample the 
number of systems with C~{\sc ii}$^*$ detections that are consistent 
with CNM and WNM are approximately equal. Most of the upper limits on the ratio, 
measured in the metallicity range $-2.0\le Z_\odot \le -1.5$, are lower than what 
would be expected from CNM gas and are consistent with WNM (or low density) gas. 
Interestingly these upper limits are lower than that seen in high latitude Galactic 
sightlines that are believed to be predominantly WNM gas. This means
that the electron density (and therefore probably the total particle density) 
in these DLAs is probably quite small.
\subsection{Systems with \h2 detection}
Systems with \h2 detections (marked as filled circles in Fig.~\ref{c2sdet}) 
have $N$(C~{\sc ii}$^*$)/$N$(C~{\sc ii}) consistent with CNM.
We compute the allowed range of n$_{\rm H}$ in these components using $N$(C~{\sc ii$^*$}),
$N$(S~{\sc ii}), $n_{\rm e}$ from the C~{\sc i} excitation and $T=T_{01}$ if available
or $T$~=~100 K (see Eq.\ref{enheq}).
The results are summarised in Column 6 of Table~\ref{tabphy1}. This Table
also gives upper limits on \nh for components without
\h2 in systems that show \h2.
It is to be remembered  that we assume [C/S] in DLAs is [C/S]$_\odot$. 
Realistically Carbon can be under-abundant by up to a factor of 2. In that case the 
density will be higher than what we quote in the table. From Table~\ref{tabphy}
and \ref{tabphy1} it is clear that for the \h2 components toward Q\,0347$-$383,
Q\,0551$-$366 and Q\,1232$+$082 the value of \nh derived from both methods agree
well. Such a comparison is not possible for the components toward Q\,0013$-$004
and Q\,0405$-$383 as C~{\sc ii$^*$} is blended in the former case and C~{\sc i}
is not detected in the latter case. In the case of Q\,1444+014 the derived hydrogen
density based on C~{\sc ii$^*$} is lower than that derived using C~{\sc i} 
fine-structure excitation. However, in this system, Ledoux et al. (2003) have found 
a 5 km/s shift between the C~{\sc i} absorption line and that of singly ionized species. 
In addition these components showing relative depletion of Si with respect to S,
it is possible that we have over estimated $N$(C~{\sc ii}). In summary, the 
$n_{\rm H}$ estimates
based on the two methods are approximately consistent with one another.
The excitation of the fine-structure levels of C~{\sc i}
and C~{\sc ii} in the components with \h2 detection are consistent with
high density and low temperature CNM gas.
\subsection{Systems without \h2 detection}
\begin{table}
\caption{Systems without \h2 detection}
{\tiny
\begin{tabular}{ccccccc}
\hline
QSO &\zabs &log x(Al~{\sc iii})& \multicolumn{4}{c}{\nh (cm$^{-3}$)}  \\
    &      &                   & CNM$^1$& WMN$^2$&Ionized$^3$& Max$^4$\\
\hline
\hline
Q\,$0058-292$& 2.671 &....   &  3.1 & 1.2 & 0.3 & $<$3  \\
Q\,$0112-306$& 2.422 &$-1.46$& 24.4 & 9.6 & 2.3 & $<$15 \\ 
Q\,$0135-273$& 2.799 &$-1.86$& 60.0 &23.4 & 5.6 & $<$5  \\
Q\,$0405-443$& 2.550 &$-1.58$&  7.3 & 1.8 & 0.4 & $<$4  \\
Q\,$0841+129$& 2.374 &$-1.16$& 11.3 & 2.8 & 0.7 & $<$22 \\
Q\,$1157+014$& 1.944 &$-1.68$& 16.3 & 4.0 & 1.0 & $<$3  \\
Q\,$1223+178$& 2.465 &$-1.35$&  9.4 & 2.3 & 0.6 & $<$2  \\
\hline
\multicolumn{6}{l}{$^1$ $T$~=~100 K and $n_{\rm e}/n_{\rm H}$ = 0.001; $^2$ $T$~=~8000 K and 
$n_{\rm e}/n_{\rm H}$=0.01}\\
\multicolumn{6}{l}{$^3$ $T$~=~10$^4$ K and $n_{\rm e}/n_{\rm H}$ = 0.1; $^4$ from \h2 equilibrium formation}
\end{tabular}
\label{c2sden}
}
\end{table}
In this Section we focus our attention on the 7 systems in our sample that
show C~{\sc ii$^*$} without \h2.           
These systems do not show detectable C~{\sc i} absorption lines
except the high-metallicity
system at $z_{\rm abs}$~=~2.1391 toward Tol~1037$-$270. 
Apart from the \zabs = 1.943 system toward Q 1157+014 that show
21 cm absorption line (Wolfe et al. 1981) there is no independent
constraint on $T$ and $n_{\rm H}$. 
The identification of C~{\sc ii$^*$} at \zabs = 2.422 toward Q\,$0112+029$ and 
\zabs = 2.799 toward Q $0135-273$ is based on C~{\sc ii$^*$}$\lambda1037$ absorption line. 
These lines are well inside the \lya forest and possible contamination by 
intervening H~{\sc i} absorption cannot be ruled out. For the rest of the systems, 
the identification and estimation of the C~{\sc ii$^*$} column density are secure.

We detect Al~{\sc iii} absorption lines in the 6 (out of 7) systems for which our spectra 
cover the expected wavelength range of the redshifted Al~{\sc iii} transitions.
We estimate the fraction of Al in Al~{\sc iii} using the observed metallicity  
and the observed $N$(Al~{\sc iii}), assuming no depletion and solar relative
abundances. Results are summarised in Table~\ref{c2sden}.
From the Table, it can be seen that 1$-$7~\% of Al is twice ionized.
Using photoionization models from ``CLOUDY'' we derive a typical 
ionization parameter $-3\le$~log~$U$~$\le-2$ if the gas originates
from a single slab irradiated by stellar spectrum with an effective 
black-body temperature of 30,000$-$40,000 K (also see Fig. 1 in Izotov et al. 2001). 
This implies that the average $n_{\rm e}/n_{\rm H}$ ratio along the line of sight 
is typically in the range 0.3 to 0.9. Thus there are enough electrons in
the cloud so that collisions with electrons are dominant in the excitation of C~{\sc ii$^*$}. 

The average density, $n_{\rm H}$, is derived using Eq.~\ref{enheq} and assuming three possible 
combinations of $T$ and $n_{\rm e}/n_{\rm H}$. The results are summarised in 
Table~\ref{c2sden}. When we use no additional constraints, the C~{\sc ii$^*$} observations 
alone are consistent with the gas having
high density and low temperature (see Column 4 of Table~\ref{c2sden}).

The last column in the table gives the upper limit on $n_{\rm H}$ that will
keep the equilibrium abundance of \h2 below our detection limit
(i.e $N$(\h2)$\le10^{14}$ cm$^{-2}$).
This value is computed using simple formation equilibrium of 
optically thin \h2 (Jura 1975)
\\
\begin{equation}
n_{\rm H} = {0.11 \beta(0) N({\rm H_2}) \over R N({\rm H I})}
\\
\label{eqh2}
\end{equation}
\noindent
with, $R$ and $\beta(0)$, respectively, the formation and 
photo-destruction rates of \h2. In the case of the ISM, $R$ 
$\simeq$ 3$\times10^{-17}$ s$^{-1}$ cm$^{-3}$ and
$\beta(0)$ $\sim 5\times 10^{-10}$ s$^{-1}$. We use the ISM value of $R$ scaled 
by the dust content measured in the systems. 
It can be seen from Table~\ref{c2sden} that for a moderate radiation field (like the 
ISM mean field) $n_{\rm H}$ derived using C~{\sc ii$^*$} for the CNM like parameters 
is usually higher than the upper limit obtained based on the \h2 
equilibrium formation. 
In addition, the expected electron density for the $T$ = 100 K gas 
(assuming a $n_{\rm e}/n_{\rm H}$ ratio as seen in CNM) is higher than 
10$^{-2}$ cm$^{-3}$. At such electron densities,
C~{\sc i} should be detectable. This is inconsistent with the non-detection
of C~{\sc i} in these systems.
This problem of CNM gas 
producing very small amount of C~{\sc i} is already recognized in the literature 
(Liszt et al. 2002; Wolfe et al. 2003).

We notice that the absence of C~{\sc i} and \h2 in these systems is consistent 
with the gas originating either from the WNM gas or from the ionized gas. As pointed out 
above, the strength of the Al~{\sc iii} absorption lines seen in these systemsare consistent 
with the gas density being less than the one expected for the CNM.

Thus, if one uses only C~{\sc ii$^*$} absorption line then the results 
are consistent with these systems originating from CNM gas. 
However the absence of \h2 and C~{\sc i} absorption lines together 
with the presence of Al~{\sc iii} following the profiles of 
singly ionized gas is inconsistent with standard CNM solutions. 
Thus most of the DLAs without \h2 are consistent with them 
originating from low density, 
high temperature and partially ionized gas.
\subsection{\zabs = 1.944 toward Q\,$1157+014$}
\begin{figure}
\flushleft{\vbox{
\psfig{figure=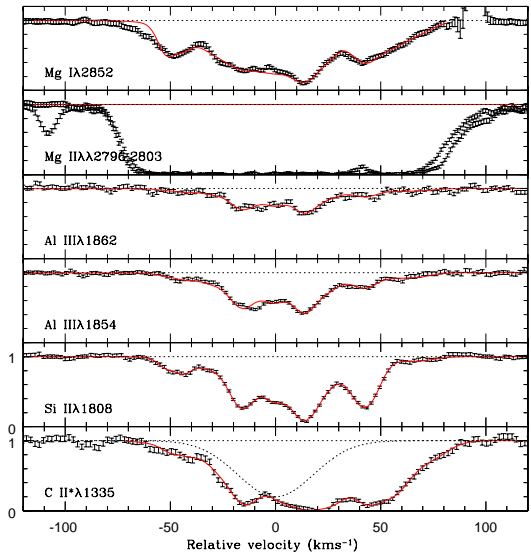,width=8.cm,height=8.cm,angle=0.}}}
\caption[]{Velocity plot for the \zabs = 1.944 system toward
Q\,$1157+014$. The dotted Gaussian profile in the bottom panel
is the reconstruction of the best fit Gaussian to the 21 cm
observations of Wolfe et al. (1981). This is just to illustrate
the velocity range over which cold hydrogen gas is detected.
}
\label{vp1157}
\end{figure}
Some of our conclusions of the previous Section can be ascertained 
in the case of the \zabs = 1.944 system toward Q\,$1157+014$ as we have additional
information on the kinetic temperature based on 21 cm absorption
(Wolfe et al. 1981). The estimated spin temperature based on the recent 
measurement of $N$(H~{\sc i}) is $T$~=~865$\pm$190~K (Kanekar \& Chengalur 2003). 
Fig.~\ref{vp1157} shows the velocity plot of selected absorption lines in this 
system. The distribution of neutral, singly ionized and doubly ionized species in  
velocity space can be visualized easily. Using a Gaussian profile, we also show, 
in the bottom panel, the velocity range over which 21~cm absorption is seen.
Whereas the UV absorption lines spread over more than 100 \kms, the 21 cm 
absorption originates from only a few of the components. There are components that 
do not possess 21cm absorption but show absorption due to UV transitions including 
that of C~{\sc ii$^*$}. As expected, C~{\sc ii$^*$} traces a wide range of 
physical conditions. 
Assuming the temperature of the gas that is producing the 21 cm absorption
feature to be 100 K (or 200 K), we can derive from the 21~cm observation the 
\hi column density in the component lying along the line of sight: 
$\sim7\times10^{20}$ cm$^{-2}$ (or 1.4$\times10^{21}$ cm$^{-2}$). This is 
approximately 15\% (30\%) of the total $N$(H~{\sc i}) measured from the damped 
Lyman-$\alpha$ line. This means that 85\% (70\%) of $N$(H~{\sc i}) along the line of sight 
is either warm or hot. 
{From the Si~{\sc ii}$\lambda$1808 profile, we notice that 
$\sim$60\% of $N$(Si~{\sc ii}) originate in the velocity space covered by the 21cm 
profile. We also notice that considerable 
fraction of Al~{\sc iii} absorption 
originate from the velocity range covered by 21 cm absorption. Thus 
warm and ionized components seem to be co-spacial with the cold 
gas responsible for the 21 cm absorption. In addition we also notice Al~{\sc iii}
components with Si~{\sc ii} and C~{\sc ii$^*$} absorption well separated from
the 21 cm component. Thus part of the C~{\sc ii$^*$} absorption seen here 
originate from the WNM or WIM. 
Therefore, the non-detection of 
\h2 in this system can be easily explained as a consequence of most of the gas being 
at high temperature (and hence low density). The absence of \h2 from the cold 21 cm absorbing 
component could just be due to the low $N$(H~{\sc i}) associated with this component and 
the relatively low dust depletion.}
\section{Discussion and conclusions}
We have studied the physical conditions in damped Lyman-$\alpha$ systems (DLAs) 
using a sample of 33 systems toward 26 QSOs acquired for a recently completed 
survey of \h2 in DLAs by Ledoux et al. (2003). We use standard techniques 
to estimate the physical conditions prevailing in the gas. In this Section, 
we discuss some of the results and related issues.
\subsection{High pressure of the \h2 gas}
Our study shows that the \h2 components in DLAs trace Cold gas (153$\pm78$ K) 
with relatively high pressure. The pressure in individual 
components (measured assuming a radiation field similar to
our Galaxy) is in the range 824$-$30,000 cm$^{-3}$ K, a large fraction 
of the components being at high pressure. 
42\%, 20\%, and 8\% of the components have pressure in excess of 
3000 cm$^{-3}$ K,  5000 cm$^{-3}$ K and 10$^4$ cm$^{-3}$ K, 
respectively. Based on the profiles of singly ionized species we 
note the \h2 components arise in gas with a wide range of molecular 
content and ionization state much like what we see in the Galactic ISM. 

This is not unexpected. Indeed,
in the framework of a galactic two-phase medium, the stable pressure range
for the gas is 460$\le {\rm P/k~(cm^{-2}~K)}\le$ 1750 (Wolfe et al. 2003). 
Clearly the pressure we derive in the \h2 components are much
higher than this. From Table 3 of Wolfire et al. (1995)
it can be seen that for a given metallicity an increase in the dust-to-gas
ratio can lead to an increase in the allowed range of pressure,
whereas an increase in the metallicity reduces the allowed range 
of pressure due to enhanced cooling. For conditions typical of DLAs,
that is for metallicities of Z = $-1.0$ and a dust-to-gas ratio ten times
smaller than in the Galaxy, the stable pressure range is 1800-13000 
cm$^{-3}$ s$^{-1}$ (Wolfire et al. 1995). 
Note that in the absence of any confining medium (or pressure equilibrium 
between different components) we expect such a high pressure gas to 
survive only for a short period of time (with a typical hydrodynamical
time-scale of 10$^6$ years). 

The pressure we infer depends very much on the intensity of the radiation 
field. A larger intensity implies and
excess of UV pumping (on top of what we assume in our analysis) 
which, if taken into account, should reduce the hydrogen density 
derived using the C~{\sc i} 
fine-structure lines. At the same time, the temperature of the gas will 
increase due to photo-heating. 
\subsection{\h2 content} 
Ledoux et al. (2003) found that approximately 13$-$20\% of DLAs show 
\h2 absorption lines
with most of the \h2 components having column densities in the 
range $16.0\le$~log~$N$(H$_2$)(cm$^{-2}$)~$\le19.0$.
In the case of the Galactic ISM, only a minor fraction of the clouds 
fall in this range. This is expected because, above 
log~$N$(\h2)$\simeq$ 16.0, self-shielding
drastically decreases the photo-dissociation rate. 
On the contrary, in the LMC and SMC, a large fraction of lines of
sight have $16.0\le$~log~$N$(H$_2$)~$\le19.0$ (see Tumlinson et al., 2002).
Thus the trend noticed in DLAs could just be a generic feature 
of gas with low dust content and metallicity. 
It is however important to remember that the molecular fraction given in 
Ledoux et al. (2003) is an average over the whole line of sight. 
The actual molecular fraction in individual components may be much larger. 
Thus the low values of $N$(\h2) that are observed could just be a 
consequence of low $N$(H~{\sc i}) in the corresponding individual components. 
Indeed, consistent models of DLAs (Srianand et al. 2005) 
require H~{\sc i} column densities much less than the total $N$(H~{\sc i}) 
measured in DLAs with \h2.
In addition, the absence of 21~cm absorption can be reconciled if the 
\h2 components have only part of the total \hi (see below).
\subsection{21 cm absorption}
{We have shown that detecting \h2 and C~{\sc i} absorption lines is an efficient 
way to trace the cold neutral gas in DLAs.
%
\hi 21 cm absorption line provides an independent  way of detecting the CNM gas in DLAs}. 
The detectability of 21 cm absorption line depends only on the amount of cold 
gas along the line of sight and the covering factor of the radio source. 
Thus, for a compact background source one can detect CNM gas with 21 cm 
without any bias from dust content or metallicity. 
At \zabs $\ge$ 2, seven systems have 
been searched for 21 cm absorption and none have been detected 
(Kanekar \& Chengalur 2003).
Assuming $T$ = 200 K for the CNM gas, these authors estimated the filling 
factor of the CNM gas to be $\le0.3$. This is consistent with what we 
derive  from our \h2 survey.
Over the redshift range covered by our survey there are three cases
for which information on 21 cm absorption and \h2 content are available.
To our surprise there seems to be no correlation between 21 cm absorption 
and \h2 absorption. 
\par
The \zabs = 2.811 system toward PKS~0528$-$255 show \h2 absorption in two 
distinct components without any corresponding 21 cm absorption (Carilli 
et al. 1996).  The upper limit on $\tau$(21cm) gives 
$N$(H~{\sc i})$\le 5\times10^{20}$ cm$^{-2}$ if the kinetic temperature is 
similar to $T$(OPR) we measure. Thus \h2 and 21 cm observations can be 
consistent with one another if no more than 20\% of the total \hi column 
density is associated with the \h2 component.
The \zabs = 1.944 system toward Q 1157$+$014 (Wolfe, Briggs \& Jauncy, 
1981; and discussion above) and \zabs = 2.04 towards PKS 0458-020 
(Briggs et al., 1989; Ge \& Bechtold, 1999) show strong 21 cm absorption 
but no C~{\sc i} or \h2 absorption. In these systems the absence of
\h2 and C~{\sc i} absorption could be either due to low 
density in the cold H~{\sc i} component or to the presence of
higher ambient radiation field.
%
\subsection{C~{\sc ii$^*$} absorption}
As pointed out before, C~{\sc ii$^*$} is detected in all the systems in which
\h2 is seen. In fact, C~{\sc ii$^*$} is also detected in the three systems
that show 21 cm absorption discussed in the previous Section. C~{\sc ii} being 
the dominant ion of Carbon in the neutral gas, it is natural to expect
C~{\sc ii$^*$} associated with both 21 cm and \h2 absorption. However,
C~{\sc ii}$^*$ is readily detected in warm neutral gas and even in ionized 
gas.
The interpretation of the origin of the C~{\sc ii$^*$} absorption is not as 
straightforward as in the case of \h2 and C~{\sc i}.
Thus, the nature of systems that do not show 21~cm absorption and/or
\h2 absorption is a matter of debate. The systems that show
C~{\sc ii$^*$} in our sample are consistent with them originating
from the CNM gas. However, the absence of C~{\sc i} and \h2
(if we take the depletion as an indicator of the presence of dust) 
and the presence
of Al~{\sc iii} are also consistent with C~{\sc ii$^*$} absorption
originating from the warm/partially ionized gas. Thus the frequency of 
occurrence
of C~{\sc ii$^*$} provides a liberal upper limit on the CNM covering
factor. A detailed investigation taking into account the constraints on
the ionization state of the gas based on N~{\sc ii}, Fe~{\sc iii}
or Al~{\sc iii} will be important to derive the exact covering factor of
CNM gas.
\subsection{Star-Formation Rate}
One of the main driver for the study of DLAs is to find out a way 
to recover the global star-formation history in a typical, moderately 
star-forming environment. The importance of DLAs in the paradigm of 
hierarchical structure formation can be appreciated from the fact that 
the mass density of baryonic matter in DLAs at $z_{\rm abs}\sim 3$ is similar 
to that of stars at present epochs (Wolfe, 1995). Studies of \lya and 
UV continuum emission  from galaxies associated with DLAs usually 
result in star formation rates (or upper limits) of a few 
M$_\odot$ yr$^{-1}$ (Fynbo et al., 1999; Bunker et al., 1999; 
Kulkarni et al., 2001). 

Wolfe et al. (2003a, 2003b) have proposed a novel idea of using the C~{\sc ii}$^*$
cooling rate to infer the SFR in DLA galaxies. The idea is that
if one assumes thermal equilibrium then the cooling rate inferred from
C~{\sc ii$^*$} should be equal to the heating rate (through UV photons,
Cosmic rays etc.,) driven by the local star-formation activity. Their detailed 
study suggests the star formation rate density of DLAs
at high-$z$ could be as high as that inferred based on Lyman break
Galaxies. We confirm the presence of C~{\sc ii$^*$}
in $\simeq$ 50\% of the DLAs in our sample. However, it is clear
from the above discussion that one needs to unveil the nature of the
partially ionized gas in order to have a handle on the heating rates.

In the local universe, star-formation is always related to molecular clouds.
If DLAs are star-forming regions then the local star-formation has
to be related to the mass of the molecular gas. Our survey shows 
that 13$-$20\% of DLAs are associated with \h2 in absorption. We have not detected 
CO in any of these systems and HD is detected in only one system (Varshalavich et 
al., 2000). Clearly the dark molecular clouds where stars form in our 
Galaxy are not seen along QSO lines of sight. The UV radiation field inferred from
the \h2 high-J excitation is similar to the Galactic mean field. Following 
Wolfe et al. (2003a,2003b) the SFR per unit comoving volume for DLAs is,
\begin{eqnarray}
\dot \rho_* &=& {An_{co}(z)<\dot\xi(z)> }\nonumber\\
            &=& f_d <\dot\xi(z)> ({A\over A_p}) {dN\over dX}
\end{eqnarray}
where, $<\dot\xi(z)>$ is the average SFR per unit area at redshift $z$
and $A,~A_{\rm p}$ and ${dN\over dX}$ are average physical cross-sectional,
respectively, area, average projected area and number density of absorbers per unit
absorption distance interval. $f_{\rm d}$ is the fraction of
DLAs in which the UV radiation field is similar to the
Galactic UV background (i.e., 0.13$-$0.20). Here we use the fact that our \h2 
sample is a randomly chosen sub-sample of the whole population of DLAs and 
the presence of \h2 is independent of $N$(H~{\sc i}).

For an Einstein-de Sitter cosmology, ${dN\over dX}=3\times10^{-5}$ 
for the mean redshift of our sample (Storrie-Lombardi \& Wolfe, 2000). Assuming 
$H_0$ = 75 \kms Mpc$^{-1}$, $A/A_{\rm p} = 2$  and 
$<\dot\xi(z)> = 4\times 10^{-3}$ M$_\odot$ yr$^{-1}$ kpc$^{-2}$
(typical for our Galaxy, see Kennicutt, 1998) we derive $\dot \rho_*\ge0.03$
at \zabs = 2.5. This crude estimate already gives half the star formation
rate density measured in Lyman break galaxies (Steidel et al., 1999).
Recently, Hirashita \& Ferrera (2005) have also arrived at similar conclusion.
Thus it is of the
utmost importance to understand the physics of the ISM in high-$z$ DLAs in order to 
derive the cosmic star-formation budget correctly.

\section*{acknowledgements}
Results presented in this work are based on observations carried out at
the European Southern Observatory (ESO) under prog. ID No. 65.P-0038, 65.O-0063, 66.A-0624,
67.A-0078, 68.A-0600 68.A-0106 and 70.A-0017 with the UVES spectrograph installed on 
the Very Large Telescope (VLT) at Cerro Paranal Observatory in Chile. 
RS and PPJ gratefully acknowledge support from the Indo-French
Centre for the Promotion of Advanced Research (Centre Franco-Indien pour
la Promotion de la Recherche Avanc\'ee) under contract No. 3004-3.
GJF acknowledges the support of the NSF 
through AST 00-71180 and NASA with grant NAG5-12020. 
GJF and RS acknowledge the support from the DST/INT/US(NSF-RP0-115)/2002.
GS would like to thank CCS, University of Kentucky for their two years of support.
The hospitality of  IUCAA is gratefully acknowledged.

\end{document}